\documentclass[aps,12pt,showpacs,preprintnumbers,nofootinbib,endfloats*]{revtex4}
\usepackage{graphicx}

\def\err#1#2{\lower2pt\hbox{ $\stackrel{\scriptstyle +#1}{\scriptstyle
-#2}$}}

\begin{document}
\title{Kaluza-Klein States of the Standard Model Gauge Bosons: \\
Constraints From High Energy Experiments}
\author{Kingman Cheung$^{a}$ and Greg Landsberg$^b$
\footnote{E-mail addresses: cheung@phys.cts.nthu.edu.tw
and landsberg@hep.brown.edu}}
\affiliation{$^a$ National Center for Theoretical Science, National Tsing
Hua University, Hsinchu, Taiwan\footnote{ The initial stage of the
work was performed at the Davis Institute for High Energy Physics at
the University of California, Davis, CA 95616, U.S.A.}\\ 
$^b$ Department of Physics, Brown University, Providence, RI 02912, USA}
\date{\today}

\begin{abstract} 
In theories with the standard model gauge bosons
propagating in TeV$^{-1}$-size extra dimensions, their Kaluza-Klein
states interact with the rest of the SM particles confined to the
3-brane.  We look for possible signals for this interaction in the present
high-energy collider data, and estimate the sensitivity offered by the
next generation of collider experiments. Based on the present data
from the LEP~2, Tevatron, and HERA experiments, we set a lower limit
on the extra dimension compactification scale $M_C > 6.8$~TeV at the
95\% confidence level (dominated by the LEP 2 results) and quote expected
sensitivities in the Tevatron Run 2 and at the LHC.
\end{abstract}

\preprint{NSC-NCTS-011026}
\preprint{FERMILAB-Pub-01/333-T}
\maketitle

\section{Introduction}

A possibility that the universe has additional compactified spatial
dimensions beyond the familiar four-dimensional space-time has been
long discussed~\cite{kk}. Advances in modern string theory, along with
the continuous attempts to solve the hierarchy problem of the standard
model (SM), have revived interests to this subject. Recently, it has
been suggested that the Planck, string, and grand unification scales
can all be significantly lower than it was previously thought, perhaps
as low as a few TeV~\cite{lowstring}. 
For example, in a viable model suggested by
Arkani-Hamed, Dimopoulos, and Dvali~\cite{arkani}, the
matter is confined to a 3-brane while gravity propagates in extra
dimensions of a sub-millimeter size.  In this model, the effective
Planck scale is as low as a TeV, thus eliminating the hierarchy
problem of the SM.
This also 
yields rich phenomenology within the reach of future collider experiments,
including production of monojets (see, e.g., \cite{GRW,Peskin,HLZ}),
modification of the Drell-Yan spectrum~(see,
e.g. \cite{GRW,HLZ,Hewett}), and even creation of mini black-holes and
string-balls~\cite{bh}. (For a brief summary of current experimental
situation, see Ref. \cite{review}.)

A more generic picture drawn in string theories is that the SM matter particles
reside on a $p$-brane ($p=\delta+3$; the space-time dimension of the
brane is then $p+1$) while gravity propagates in the entire
ten-dimensional bulk.  The compactification of the $\delta$ dimensions
occurs {\it internally} within the brane, while the remaining
$(6-\delta)$ dimensions are compactified {\it transverse} to the
brane.  Various phenomenology arises, depending on the relative
magnitude of the two compactification scales, the string scale, and
the Planck scale. 
The model of Arkani-Hamed, Dimopoulos, and Dvali~\cite{arkani} is
a specific example with $\delta=0$.

Another interesting model was also proposed \cite{keith,theory}, in
which matter resides on a $p$-brane ($p>3$), with chiral fermions
confined to the ordinary three-dimensional world internal to the
$p$-brane and the SM gauge bosons also propagating in the extra $\delta >
0$ dimensions internal to the $p$-brane. (Gravity in the bulk is not
of direct concern in this model.) It was shown \cite{keith} that in
this scenario it is possible to achieve the gauge coupling
unification at a scale much lower than the usual GUT scale, due to a
much faster power-law running of the couplings at the scales above
the compactification scale of the extra dimensions.
The SM gauge bosons that propagate in the extra dimensions
compactified on $S^1/Z_2$, in the four-dimensional point of view,
are equivalent to towers of Kaluza-Klein (KK) states with masses
$M_n = \sqrt{M_0^2 + n^2/R^2}$ ($n=1,2, ...$), where $R=M_C^{-1}$ 
is the size of the compact dimension, $M_C$ is the corresponding 
compactification scale, and $M_0$ is the mass of the corresponding SM 
gauge boson. 

There are two important consequences of the existence of 
the KK states of the gauge bosons in collider phenomenology. (i) Since 
the entire tower of KK states have the same quantum numbers as their
zeroth-state gauge boson, this gives rise to mixings among the
zeroth (the SM gauge boson) and the $n$th-modes ($n=1,2,3,...$) 
of the $W$ and $Z$ bosons.
(The zero mass of the photon is protected by the U(1)$_{\rm EM}$
symmetry of the SM.) (ii) In addition to direct production and virtual
exchanges of the zeroth-state gauge bosons, both direct production and
virtual effects of the KK states of the $W,Z,\gamma$, and $g$ bosons
would become possible at high energies.

In this paper, we study the effects of virtual exchanges of the KK
states of the $W,Z,\gamma$, and $g$ bosons in high energy collider
processes.  While the effects on the low-energy precision measurements
have been studied in detail
\cite{nath,rizzo,casa,Strumia,carone,del,CC} (we shall briefly
summarize their findings in a later section), their high-energy
counterparts have not been systematically studied yet. We attempt to
bridge this gap by analyzing all the available high-energy collider
data including the dilepton, dijet, and top-pair production at the
Tevatron; neutral and charged-current deep-inelastic scattering at
HERA; and the precision observables in leptonic and hadronic
production at LEP~2.

We fit the observables in the above processes to the sum of the SM
prediction and the contribution from the KK states of the SM gauge
bosons. In all cases, the data do not require the presence of the
KK excitations, which is then translated to the limits on the
compactification scale $M_C$. The fit to the combined data set yields
a 95\% C.L. lower limit on $M_C$ of 6.8 TeV, which is substantially
higher than that obtained using only electroweak precision measurements.
In addition, we also estimate the expected reach on $M_C$ in
Run 2 of the Fermilab Tevatron and at the LHC, using dilepton
production.

The organization of this paper is as follows: In the next section, we
describe the Lagrangian for the model~\cite{theory}, which has one
extra dimension.  In Sec.~\ref{sec:low-energy} we briefly summarize the existing 
constraints from precision measurements. In Sec.~\ref{sec:APV}, we briefly 
discuss the effects of the KK states of the $Z$ boson on the atomic parity 
violation (APV) measurements. In Sec.~\ref{sec:data} we describe the high energy 
data sets that we used in this analysis. In Sec.~\ref{sec:results}, we present our 
results on the fits and limits. In Sec.~\ref{sec:sensitivity}, we estimate the 
sensitivity in Run 2 of the Tevatron and at the LHC.
A collection of data sets that we used in our analysis is placed 
in the appendix.

\section{Interactions of the Kaluza-Klein States}

In what follows, we use the formalism of Ref. \cite{theory}, based on
an extension of the SM to five dimensions, with the fifth dimension,
$x^5$, compactified on the segment $S^1/Z_2$ (a circle of radius $R$
with the identification $x^5 \to -x^5$). This segment has the length
of $\pi R$. Two 3-branes reside at the fixed points $x^5=0$ and
$x^5=\pi R$. The SM gauge boson fields propagate in the 5D-bulk, while
the SM fermions are confined to the 3-brane located at $x^5=0$. 
The Higgs sector consists of two Higgs doublets, $\phi_1$ and $\phi_2$ 
(with the ratio of vacuum expectation values $v_2/v_1 \equiv \tan\beta$), 
which live in the bulk and on the SM brane, respectively.

The 5D Lagrangian is given by
$$
	{\cal L}_5 = - \frac{1}{4g_5} F^2_{MN} + \left| D_M
	\phi_1 \right |^2 +\left( i \bar\psi \sigma^\mu D_\mu \psi +
	\left|D_\mu \phi_2 \right |^2
	\right )\delta(x^5) \;,
$$ 
where $D_M=\partial_M + iV_M$, $M=(\mu,5)=(1,...,5)$,
and $g_5$ is the 5D gauge coupling for the gauge boson $V$.
Compactifying the fifth dimension on $S^1/Z_2$ with the expansion
$$
	\Phi(x^\mu, x^5) = \sum_{n=0}^{\infty} \cos \left(\frac{n x^5}{R}
	\right)
	\Phi^{(n)}(x^\mu) \;,
$$
the 4D Lagrangian becomes~\cite{theory}
\begin{eqnarray}
\label{l4} 
	{\cal L}_4 &=& \sum_{n=0}^{\infty} \left[ -\frac{1}{4}
	F^{(n)2}_{\mu\nu} + \frac{1}{2} \left( \frac{n^2}{R} + 2 g^2
	|\phi_1|^2 \right ) V^{(n)}_\mu V^{(n)\mu} \right ] + g^2 |\phi_2|^2
	\left( V_\mu^{(0)} + \sqrt{2} \sum_{n=1}^\infty V^{(n)}_\mu \right )^2
	\nonumber \\ && + i \bar \psi \sigma^\mu \left[
	\partial_\mu + igV_\mu^{(0)} + i g \sqrt{2} \sum_{n=1}^\infty
	V_\mu^{(n)} \right ] \psi + ... \;,
\end{eqnarray} 
where $g = g_5/\sqrt{\pi R}$ is the 4D gauge coupling for the gauge boson $V$.

In the case of SU(2)$_L \times$ U(1)$_Y$ symmetry, the charged-current
(CC) and neutral-current (NC) interactions after compactifying the
fifth dimension are given by \cite{casa}:
\begin{eqnarray}
\label{cc} 
	{\cal L}^{\rm CC} &=& \frac{g^2 v^2}{8} \biggr [ W_1^2 +
	\cos^2\beta
	\sum_{n=1}^{\infty}( W_1^{(n)})^2 + 2\sqrt{2} \sin^2\beta W_1
	\sum_{n=1}^{\infty} W_1^{(n)} + 2 \sin^2\beta \left(
	\sum_{n=1}^{\infty} W_1^{(n)} \right )^2 \biggr ] \nonumber \\ &+&
	\frac{1}{2} \sum_{n=1}^{\infty} n^2 M_C^2 (W_1^{(n)})^2 - {g} (W_1^\mu
	+ \sqrt{2} \sum_{n=1}^{\infty} W_1^{(n)\mu} ) J_\mu^1 +(1 \to 2) \;,
	\\
\label{nc} 
	{\cal L}^{\rm NC} &=& \frac{{g}v^2}{8 {c}_\theta^2} \biggr
	[ Z^2 + \cos^2\beta \sum_{n=1}^{\infty}( Z^{(n)})^2 + 2\sqrt{2}
	\sin^2\beta Z
	\sum_{n=1}^{\infty} Z^{(n)} + 2 \sin^2\beta \left (\sum_{n=1}^{\infty}
	Z^{(n)} \right )^2 \nonumber \\ &+& \frac{1}{2} \sum_{n=1}^{\infty}
	n^2 M_C^2 \biggr[ (Z^{(n)} )^2 + (A^{(n)})^2 \biggr] \nonumber \\ &-&
	\frac{{e}}{ {s}_\theta {c}_\theta } \left ( Z^\mu + \sqrt{2}
	\sum_{n=1}^{\infty} Z^{(n)\mu} \right ) J_\mu^Z - {e} \left ( A^\mu +
	\sqrt{2} \sum_{n=1}^{\infty} A^{(n)\mu} \right ) J_\mu^{\rm em} \;,
\end{eqnarray} 
where the fermion currents are:
\begin{eqnarray} 
	J_\mu^{1,2} &=& \bar \psi_L \gamma_\mu
	\left(\frac{\tau_{1,2}}{2} \right )
	\psi_L \;, \nonumber \\ 
	J_\mu^Z &=& \bar \psi \gamma_\mu ( g_v - \gamma^5 g_a )
	\psi \;, \nonumber \\ 
	J_\mu^{\rm em} &=& \bar \psi \gamma_\mu Q_\psi \psi \;,\nonumber
\end{eqnarray} 
and $\langle \phi_1 \rangle = v \cos\beta, \langle
\phi_2 \rangle=v\sin\beta$; $g$ and $g'$ are the gauge couplings
of the SU(2)$_L$ and U(1)$_Y$, respectively; $g_v = T_{3L}/2 -
s_\theta^2 Q$ and $g_a = T_{3L}/2$. Here, we used the following short-hand
notations: $s_\theta \equiv \sin\theta_W$ and $c_\theta \equiv
\cos\theta_W$, where $\theta_W$ is the weak-mixing
angle. The tree-level (non-physical) $W$ and $Z$ masses are $M_W =
gv/2$ and $M_Z = M_W /c_\theta$. Since the compactification scale
$M_C$ is expected to be in the TeV range, we therefore ignore 
in the above equations the mass of the zeroth-state gauge boson 
in the expression for the mass of the $n$-th KK excitation: 
$M_n = \sqrt{M_0^2 + n^2 M_C^2} \approx n M_C$, $n=1,2,...$.

Using the above Lagrangians we can describe the two major effects of
the KK states: mixing with the SM gauge bosons and virtual exchanges
in high-energy interactions.

\subsection{Mixing with the SM Gauge Bosons}

The first few terms in the Eqs. (\ref{cc}) and (\ref{nc}) imply the
existence of mixings among the SM boson ($V$) and its KK excitations
($V^{(1)}$, $V^{(2)},\; ...$) where $V=W,Z$. There is no mixing for the 
$A^\mu$ fields because of the U(1)$_{\rm EM}$ symmetry. These mixings 
modify the electroweak observables (similar to the mixing between the 
$Z$ and $Z'$). The SM weak eigenstate of the $Z$-boson, $Z^{(0)}$, 
mixes with its excited KK states $Z^{(n)}$ ($n=1,2,...$) via a series 
of mixing angles, which depend on the masses of $Z^{(n)}, n=0,1,...$ 
and on the angle 
$\beta$. The $Z$ boson studied at LEP~1 is then the lowest mass 
eigenstate after mixing. The couplings of the $Z^{(0)}$ to fermions 
are also modified through the mixing angles. The observables at LEP~1 
can place strong constraints on the mixing, and thus on the 
compactification scale $M_C$. Similarly, the properties of the $W$ boson 
are also modified. However, so far the mass and couplings of the $W$ 
are not measured as precisely as the $Z$ observables, so the constraints 
on $M_C$ coming from the $W$ are weaker than those from the $Z$.

The effects on electroweak precision measurements have been previously
studied~\cite{nath,rizzo,casa, Strumia,carone,del,CC}; we will
summarize their results in the next section.

\subsection{Virtual Exchanges}

If the available energy is higher than the compactification scale the
on-shell production of the Kaluza-Klein excitations of the gauge
bosons can be observed~\cite{rizzo1}. However, for the present
collider energies only indirect effects can be seen, as the 
compactification scale is believed to be at least a few TeV. These
indirect effects are due to virtual exchange of the KK-states.

When considering these virtual exchanges, we ignore a slight
modification of the coupling constants to fermions due to the mixings
among the KK states and so we use Eqs. (\ref{cc}) and (\ref{nc}) without 
the mixings\footnote{Since $M_C >> M_Z$, the mixings are very small.  
Furthermore, they completely vanish for  $\beta=0$.}. 
This implies that any Feynman 
diagram which has an exchange of a $W$, $Z$, $\gamma$, or $g$ will be 
replicated for every corresponding KK state with the masses $n M_C,$ 
where $n=1,2,...$.  Note that the coupling constant of the KK states 
to fermions is a factor of $\sqrt{2}$ larger than that for the 
corresponding SM gauge boson, due to the normalization of the KK 
excitations.

It has been shown in Ref.~\cite{keith} that in the presence of the KK
states of gauge bosons in the bulk, the renormalization-group
evolution of the gauge couplings changes from the normal logarithmic
running to a power running for energy scales above $M_C$.  However,
the energy scale of the processes that we consider in this paper is
well below $M_C$. Consequently, the running of gauge couplings is the
same as the normal logarithmic running in the
SM~\cite{keith}. Besides, we are not concerned about the additional
real scalars transforming in the adjoint of each gauge group that are
required to give masses to the gauge bosons~\cite{keith}. This is
because the scalars usually couple to light fermions via very small
Yukawa couplings.

We start with Drell-Yan production of a pair of leptons. The
amplitude squared for $q\bar q \to \ell^+ \ell^-$ or $\ell^+
\ell^- \to q\bar q $ (without averaging over the initial spins or
colors) is given by:
$$
	\sum |{\cal M}|^2 = 4 u^2 \left( |M^{\ell q}_{\rm LL}(s)|^2 + |M^{\ell
	q}_{\rm RR}(s)|^2 \right ) + 4 t^2 \left( |M^{\ell q}_{\rm LR}(s)|^2 +
	|M^{\ell q}_{\rm RL}(s)|^2 \right ) \;,
$$
where
\begin{eqnarray} 
	M^{\ell q}_{\alpha\beta}(s) &=& e^2 \Biggr \{\frac{Q_\ell Q_q}{s} +
	\frac{g_\alpha^\ell g_\beta^q}{\sin^2\theta_W \cos^2 \theta_{\rm
	w} } \; 
	\frac{1}{s - M_Z^2 } \nonumber \\ &+& 2 \sum_{n=1}^\infty \biggr [
	\frac{Q_\ell Q_q}{s - n^2 M_C^2} + \frac{g_\alpha^\ell
	g_\beta^q}{\sin^2\theta_W \cos^2 \theta_W } \; \frac{1}{s
	- n^2 M_C^2 } \biggr ] \; \Biggr \} \;.\nonumber
\end{eqnarray} 
Here $s,t,u$ are the usual Mandelstam variables, $g_L^f
= T_{3f} - Q_f \sin^2\theta_W$, $g_R^f = - Q_f \sin^2\theta_{\rm
w}$, and $Q_f$ is the electric charge of the fermion $f$ in units
of proton charge.

If the compactification scale $M_C \gg \sqrt{s}, \sqrt{|t|}, \sqrt{|u|}$, 
the above can further be simplified to:
\begin{equation}
\label{mab} 
	M^{\ell q}_{\alpha\beta}(s) = e^2 \Biggr \{ \frac{Q_\ell Q_q}{s} +
	\frac{g_\alpha^\ell g_\beta^q}{\sin^2\theta_W \cos^2 \theta_W } \;
	\frac{1}{s - M_Z^2 } - \left( Q_\ell Q_q + \frac{g_\alpha^\ell
	g_\beta^q} {\sin^2\theta_W \cos^2 \theta_W } \right ) \;
	\frac{\pi^2}{3 M_C^2 } \; \Biggr \} \;.
\end{equation}

Based on the above formula the amplitude squared for deep-inelastic
scattering at HERA can be obtained by a simple interchange of the
Mandelstam variables. In the later section, we will derive the
expressions for specific observables used in our analysis.

\section{Review of the Low-Energy Constraints}
\label{sec:low-energy}

The effects of KK excitations in the low-energy limit can be included
by eliminating their fields using equations of motion following from 
the Lagrangians given by Eqs. (\ref{cc}) and (\ref{nc})~\cite{casa}:
\begin{eqnarray} 
	W_{1,2}^{(n)} &=& - \sqrt{2} \frac{ \sin^2\beta
	M_W^2}{n^2 M_C^2} W_{1,2} + \sqrt{2}\frac{g}{n^2 M_C^2} J^{1,2} 
	+ O(1/M_C^4) \;, \nonumber \\ 
	Z^{(n)} &=& - \sqrt{2} \frac{ \sin^2\beta M_Z^2}{n^2 M_C^2} Z +
	\sqrt{2}\frac{e}{s_\theta c_\theta} \frac{1}{n^2 M_C^2} J^Z + 
	O(1/M_C^4) \;, \nonumber \\ 
 	A^{(n)} &=& \sqrt{2} \frac{e}{n^2 M_C^2} J^{\rm em} + O(1/M_C^4) 
	\nonumber \;.
\end{eqnarray} 
Substituting back into Eqs. (\ref{cc}) and (\ref{nc})
we obtain the physical $W$ and $Z$ masses and the interaction
Lagrangian~\cite{casa}:
\begin{eqnarray} 
	M_W^2 &=& M_W^2 ( 1 - c^2_\theta \sin_\beta^4 X ) \;, \nonumber \\ 
	M_Z^2 &=& M_Z^2 ( 1 - \sin_\beta^4 X ) \;, \nonumber \\ 
\label{cc-low} 
	{\cal L}^{\rm CC}_{\rm int} &=& - g J^1_\mu W^{1\mu} (1-\sin^2\beta
	c^2_\theta X ) - \frac{g^2}{2M_Z^2} X J^1_\mu J^{1\mu} + (1\to 2) \;,
	\\ 
\label{nc-low} 
	{\cal L}^{\rm NC}_{\rm int} &=& - \frac{e}{s_\theta
	c_\theta} J^Z_\mu Z^\mu (1-\sin^2\beta X ) - \frac{e^2}{2 s_\theta^2
	c_\theta^2 M_Z^2} X J^Z_\mu J^{Z\mu} \nonumber \\ 
	&& - \; e J_\mu^{\rm
	em} A^{\mu} - \frac{e^2}{2 M_Z^2} X J^{\rm em}_\mu J^{{\rm em} \mu}
	\;, \\ 
	X &=& \frac{\pi^2 M_Z^2}{3 M_C^2} \;.\nonumber
\end{eqnarray}

The above low-energy Lagrangian already includes the effects of
gauge-boson mixings and of virtual exchange of the KK states and thus
can be used to calculate the precision observables.  We illustrate
this with a few examples. Using Eq.~(\ref{cc-low}) we can calculate
$G_F$ in muon decay:
$$
	G_F = \frac{\sqrt{2} g^2}{8 M_W^2} ( 1+ c_\theta^2 X)
	(1-2 \sin^2 \beta c_\theta^2 X ).
$$
Analogously, the partial width of the $Z$ boson into a
pair of fermions can be calculated using Eq. (\ref{nc-low}):
$$
	\Gamma(Z\to f\bar f) = \frac{N_c M_Z}{12\pi} \frac{e^2}{s_\theta^2
	c_\theta^2} \, ( 1- 2 \sin^2\beta X) \,( g_v^2 + g_a^2) \;,
$$ 
where $N_c=1(3)$ for leptons (quarks). Other quantities
can be derived similarly.

In the following, we summarize the results presented in
Refs. \cite{nath,rizzo,casa, Strumia,carone,del,CC}. Nath and
Yamaguchi~\cite{nath} used data on $G_F$, $M_W$, and $M_Z$ and set
the lower limit on $M_C \agt 1.6$~TeV. Carone~\cite{carone} studied a
number of precision observables, such as $G_F$, $\rho$, $Q_W$, leptonic
and hadronic widths of the $Z$. The most stringent constraint on $M_C$
comes from the hadronic width of the $Z$: $M_C >3.85$ TeV. Strumia
\cite{Strumia} obtained a limit $M_C > 3.4 - 4.3$ TeV from a set of
electroweak precision observables. Casalbuoni {\it et al.} 
\cite{casa} used the complete set of precision measurements, as well as
$Q_W$ and $R_\nu$'s from $\nu$-N scattering experiments, and obtained a
limit $M_C > 3.6$ TeV. Rizzo and Wells \cite{rizzo} used the
same set of data as the previous authors and obtained a limit
$M_C > 3.8$ TeV. Cornet {\it et al.} \cite{CC} used the unitarity of
the CKM matrix elements and were able to obtain a limit $M_C > 3.3$
TeV. Delgado {\it et al.} \cite{del} studied a scenario in which
quarks of different families are separated in the extra spatial
dimension and set the limit $M_C > 5$ TeV in this scenario.

\section{Atomic Parity Violation}
\label{sec:APV}

The 1999 atomic parity violation (APV) measurement on cesium
\cite{apv} has drawn a great deal of attention because the data showed
a $2.3\sigma$ deviation from the SM prediction. Several explanations
involving physics beyond the SM, such as extra $Z$ bosons
\cite{extra-z} and leptoquarks \cite{bc}, have been suggested. Later,
however, the theoretical calculations used in the analysis had been
questioned and new calculations appeared since~\cite{atom}. As a result,
data now agree with the SM prediction~\cite{langacker}:
$$
	\Delta Q_W \equiv Q_W({\rm Cs}) - Q_W^{\rm SM}({\rm Cs}) = 0.44 \pm
	0.44 \;.
$$
The KK states of the $Z$ boson act similarly to a large
number of extra $Z$ bosons with the same chiral couplings as the SM
$Z$ boson. These KK states result in a non-zero $\Delta Q_W$.

The change in $Q_W$ due to the KK states of the $Z$ in terms of chiral 
couplings is given by \cite{bc}:
\begin{eqnarray}
	\Delta Q_W &=& (-11.4\;{\rm TeV}^2) \left( \frac{-e^2}{\sin^2
	\theta_W
	\cos^2 \theta_W} \right ) ( -g_L^e + g_R^e) ( g_L^u + g_R^u)
	\eta
	\nonumber \\ 
	&+& (-12.8\;{\rm TeV}^2) \left( \frac{-e^2}{\sin^2
	\theta_W
	\cos^2 \theta_W} \right ) ( -g_L^e + g_R^e) ( g_L^d + g_R^d)
	\eta
	\nonumber \\ 
	&\approx & (-0.6\;{\rm TeV}^2) \; \eta \;, \label{eq:apv}
\end{eqnarray} 
where $\eta = \pi^2/(3M_C^2)$, $g_{L,R}^f = T_{3f} -Q_f\sin^2\theta_W$,
and $\theta_W$ is the weak mixing angle. As seen from Eq.~(\ref{eq:apv}),
the KK states with the same chiral couplings as the SM $Z$ boson give 
negative contributions to $Q_W$'s, and therefore are disfavored by the data.

\section{High Energy Processes and Data Sets}
\label{sec:data}

Before describing the data sets used in our analysis, let us first
specify certain important aspects of the analysis technique. Since the
next-to-leading order (NLO) calculations do not exist for the new
interactions yet, we use leading order (LO) calculations for 
contributions both from the SM and from new interactions, for 
consistency. However, in many cases, e.g. in the analysis of precision 
electroweak parameters, it is important to use the best available 
calculations of their SM values, as in many cases data is sensitive to 
the next-to-leading and sometimes even to higher-order corrections. Therefore, 
we normalize our leading order calculations to either the best
calculations available, or to the low-$Q^2$ region of the data set, where
the contribution from the KK states is expected to be vanishing. This
is equivalent to introducing a $Q^2$-dependent $K$-factor and using
the same $K$-factor for both the SM  contribution and the effects of 
the KK resonances, which is well justified by the similarity between these 
extra resonances and the corresponding ground-state gauge boson. The 
details of this procedure for each data set are given in the corresponding
section. Wherever parton distribution functions (PDFs) are needed, we
use the CTEQ5L (leading order fit) set \cite{cteq5}. The reason to use the 
LO PDF set is that LO PDFs are extracted using LO cross section calculations, 
thus making them more consistent with our approach.

\subsection{HERA Neutral and Charged Current Data}

ZEUS \cite{zeus} and H1 \cite{H1} have published results on the
neutral-current (NC) and charged-current (CC) deep-inelastic
scattering (DIS) in $e^+ p$ collisions at $\sqrt{s} \approx 300$
GeV. The data sets collected by H1 and ZEUS correspond to an
integrated luminosities of 35.6 and 47.7 pb$^{-1}$, respectively. H1
\cite{H1} has also published NC and CC analysis for the most recent data
collected in $e^- p$ collisions at $\sqrt{s} \approx 320 $ GeV with an
integrated luminosity of $16.4$ pb$^{-1}$.

We used single-differential cross sections $d\sigma/d Q^2$ presented by 
ZEUS \cite{zeus} and double-differential cross sections $d^2 \sigma/dx dQ^2$
published by H1 \cite{H1}. The double-differential cross section for NC DIS
in the $e^+ p$ collisions, including the effect of the KK states of the
$\gamma$ and $Z$, is given by
\begin{eqnarray}
	&&\frac{d^2\sigma}{dx dQ^2} (e^+p\to e^+X) = \nonumber \\ 
	&&  \frac{1}{16\pi}\; \Biggr \{
	\sum_q f_q(x) \,\biggr [ (1-y)^2 ( |M^{eq}_{LL}(t)|^2 +
	|M^{eq}_{RR}(t)|^2) + |M^{eq}_{LR}(t)|^2 + |M^{eq}_{RL}(t)|^2 \biggr ]
\label{nccc}
	\\ 
	&& + \sum_{\bar q} f_{\bar q} (x) \, \biggr [ |M^{eq}_{LL}(t)|^2 +
	|M^{eq}_{RR}(t)|^2 + (1-y)^2 (|M^{eq}_{LR}(t)|^2 + |M^{eq}_{RL}(t)|^2)
	\biggr ] \; \Biggr \} \;, \nonumber 
\end{eqnarray} 
where $Q^2 = sxy$ is the square of the momentum transfer and 
$f_{q/\bar q}(x)$ are parton distribution functions. The
reduced amplitudes $M_{\alpha\beta}^{eq}$ are given by
Eq. (\ref{mab}). The double differential cross section for CC DIS,
including the effect of KK states of $W$, can be written as
\begin{equation}
	\label{cccc}
	\frac{d^2\sigma}{dx dQ^2} (e^+p\to \bar \nu X) =  \frac{g^4}{64\pi}
	\; \left | \frac{1}{-Q^2-M_W^2} - \frac{\pi^2}{3 M_C^2} \right |^2
	\; \biggr [ (1-y)^2 ( d(x) + s(x)) + \bar u(x) + \bar c(x) \biggr ]
	\;,
\end{equation} 
where $d(x),s(x),\bar u(x), \bar c(x)$ are the parton
distribution functions.  The single differential cross section
$d\sigma/d Q^2$ is obtained from the above equations by integrating
over $x$. The cross section in the $e^- p$ collisions can be
obtained by interchanging $( LL \leftrightarrow LR, RR \leftrightarrow
RL)$ in Eq. (\ref{nccc}) and by interchanging $(q(x) \leftrightarrow
\bar q(x) )$ in Eq. (\ref{cccc}).

We normalize the tree-level SM cross section to that measured in the low-$Q^2$ 
($Q^2 \alt 2000$ GeV$^2$) data by a scale factor 
$C$ ($C$ is very close to 1 numerically). The cross 
section $\sigma$ used in the fitting procedure is given by
\begin{equation}
	\label{sign}
	\sigma = C \left( \sigma_{\rm SM} + \sigma_{\rm interf} +
	\sigma_{\rm KK} \right )\; ,
\end{equation} 
where $\sigma_{\rm interf}$ is the interference term between the SM and 
the KK states and $\sigma_{\rm KK}$ is the cross section due to the 
KK-state interactions only.

\subsection{Drell-Yan Production at the Tevatron}

Both CDF \cite{dy-cdf} and D\O\ \cite{dy-d0} measured the differential
cross section $d\sigma/dM_{\ell\ell}$ for Drell-Yan production, where
$M_{\ell\ell}$ is the invariant mass of the lepton pair. (CDF analyzed
data in both the electron and muon channels; D\O\ analyzed only the
electron channel.)

The differential cross section, including the contributions from the
KK states of the photon and $Z$, is given by
$$
	\frac{d^2\sigma}{dM_{\ell\ell} dy} = K \frac{M_{\ell\ell}^3}{72\pi s}
	\;
	\sum_q f_q(x_1) f_{\bar q}(x_2)\; \left( |M^{eq}_{LL}(\hat s)|^2 +
	|M^{eq}_{LR}(\hat s)|^2 + |M^{eq}_{RL}(\hat s)|^2 + |M^{eq}_{RR}(\hat
	s)|^2 \right ) \;,
$$
where $M_{\alpha\beta}^{eq}$ is given by Eq. (\ref{mab}), 
$\hat s=M^2_{\ell\ell}$, $\sqrt{s}$ is the center-of-mass energy in the 
$p\bar p$ collisions, $M_{\ell\ell}$ and $y$ are the invariant mass 
and the rapidity of the lepton pair, respectively, and $x_{1,2} = 
\frac{M_{\ell\ell}}{\sqrt s} e^{\pm y}$. 
The variable $y$ is integrated numerically to obtain the invariant
mass spectrum.   The QCD $K$-factor is given by $K= 1+\frac{\alpha_s(\hat s)}
{2 \pi} \frac{4}{3} (1 + \frac{4\pi^2}{3})$.
We scale this tree-level SM cross section 
by normalizing it to the $Z$-peak cross section measured with the data. 
The cross section used in the fitting procedure is then obtained 
similarly to that in Eq. (\ref{sign}).

\subsection{LEP~2 Data}

We analyze LEP~2 observables sensitive to the effects of the KK states
of the photon and $Z$, including hadronic and leptonic cross sections 
and forward-backward asymmetries. The LEP Electroweak Working
Group combined the $q \bar q, \mu^+ \mu^-$, and $\tau^+ \tau^-$ data 
from all four LEP collaborations \cite{lepew} for the machine energies
between 130 and 202 GeV. We use the following quantities in our
analysis: (i) total hadronic cross sections; (ii) total $\mu^+ \mu^-,
\tau^+ \tau^-$ cross sections; (iii) forward-backward asymmetries in
the $\mu$ and $\tau$ channels; and (iv) ratio of $b$-quark and
$c$-quark production to the total hadronic cross section, $R_b$ and $R_c$. We 
take into account the correlations of the data points in
each data set as given by~\cite{lepew}.

For other channels we use various data sets from individual experiments.
They are \cite{aleph,delphi,L3,opal}: (i) Bhabha scattering cross section 
$\sigma(e^+ e^- \to e^+ e^-)$; (ii) angular distribution or 
forward-backward asymmetry in hadroproduction $e^+ e^- \to q \bar q$;
(iii) angular distribution or forward-backward asymmetry in the $e^+
e^-$, $\mu^+ \mu^-$, and $\tau^+ \tau^-$ production.

The angular distribution for $e^- e^+ \to f \bar f\; (f=q,e,\mu,\tau)$
is given by
\begin{eqnarray}
	&&\frac{d\sigma}{d\cos\theta} = \nonumber \\
	&&\frac{N_f s}{128 \pi} \, \Bigg\{
	(1+\cos\theta)^2 \, \left( |M^{ef}_{LL}(s)|^2 + |M^{ef}_{RR}(s) |^2
	\right ) +(1-\cos\theta)^2 \, \left( |M^{ef}_{LR}(s)|^2 +
	|M^{ef}_{RL}(s) |^2 \right ) \nonumber \\
	&& +\delta_{ef} \biggr[(1+\cos\theta)^2 \,\left( |M^{ef}_{LL}(s)+
	M^{ef}_{LL}(t)|^2 + |M^{ef}_{RR}(s)+M^{ef}_{RR}(t)|^2 - 
	|M^{ef}_{LL}(s)|^2 - |M^{ef}_{RR}(s) |^2 \right ) \nonumber \\ 
	&& +4 \left( |M^{ef}_{LR}(t)|^2 + |M^{ef}_{RL}(t) |^2 \right
	) \biggr] \; \Biggr \} \;,\nonumber
\end{eqnarray} 
where $N_f=1$ (3) for $\ell$ $(q)$, and $M_{\alpha\beta}^{ef}$ is given 
by Eq. (\ref{mab}).  The additional terms for $f=e$ arise from the 
$t$,$u$-channel exchange diagrams.

To minimize the uncertainties from higher-order corrections, we
normalize the tree-level SM calculations to the NLO cross section,
quoted in the corresponding experimental papers. We then scale our
tree-level results, including contributions from the KK states of the
$Z$ and $\gamma$, with this normalization factor, similar to
Eq. (\ref{sign}). When fitting angular distribution, we fit to the
shape only, and treat the normalization as a free parameter of the
fit.

\subsection{Kaluza-Klein states of the Gluon in the Dijet Production at
the Tevatron}

Since the gauge bosons propagate in extra dimensions, the Kaluza-Klein 
momentum conservation applies at their self-coupling vertices. Because 
of this conservation, the triple interaction vertex with two gluons on 
the SM 3-brane and one KK state of the gluon in the bulk vanishes. 
(However, the quartic vertex with two gluons on the SM 3-brane and two 
gluon KK states in the bulk does exist.) That is why the Lagrangian in 
Eq. (\ref{l4}) only has the interactions of KK states of the gluon with 
fermions, but not with gluons. (Furthermore, if we treated the trilinear 
interaction between the gluons and the KK states of the gluon the same 
as the SM triple-gluon interaction, the gauge invariance would be 
violated at the order of $(1/M_C^2)$.)

The formulas for dijet production, including the contributions 
from KK states of the gluon (summed over the final-state and averaged 
over the initial-state helicities and colors), are:
\begin{eqnarray}
	\overline{\sum} \left| {\cal M} (qq' \to q q')\right |^2 &=&
	\frac{4}{9} g_s^4 ( {\hat s}^2 + {\hat u}^2 ) \left( \frac{1}{\hat t}
	- \frac{\pi^2}{3 M_C^2} \right )^2 \;, \nonumber \\
	\overline{\sum} \left| {\cal M} (qq \to qq)\right |^2 &=& g_s^4
	\Biggr[
	\frac{4}{9} ( {\hat s}^2 + {\hat u}^2 ) \left( \frac{1}{\hat t} -
	\frac{\pi^2}{3 M_C^2} \right )^2 + \frac{4}{9} ( {\hat s}^2 + {\hat
	t}^2 ) \left( \frac{1}{\hat u} - \frac{\pi^2}{3 M_C^2} \right )^2
	\nonumber \\ 
	&& - \frac{8}{27} {\hat s}^2 \left( \frac{1}{t} - \frac{\pi^2}{3 M_C^2}
	\right ) \left( \frac{1}{\hat u} - \frac{\pi^2}{3 M_C^2} \right )
	\Biggr ]\;, \nonumber \\
	\overline{\sum} \left| {\cal M} (q\bar q \to q' \bar q')\right |^2 &=&
	\frac{4}{9} g_s^4 ( {\hat t}^2 + {\hat u}^2 ) \left( \frac{1}{\hat s}
	- \frac{\pi^2}{3 M_C^2} \right )^2 \;, \nonumber \\
	\overline{\sum} \left| {\cal M} (q\bar q \to q\bar q)\right |^2 &=&
	g_s^4 \Biggr[
	\frac{4}{9} ( {\hat t}^2 + {\hat u}^2 ) \left( \frac{1}{\hat s} -
	\frac{\pi^2}{3 M_C^2} \right )^2 + \frac{4}{9} ( {\hat s}^2 + {\hat
	u}^2 ) \left( \frac{1}{\hat t} - \frac{\pi^2}{3 M_C^2} \right )^2
	\nonumber \\ 
	&& - \frac{8}{27} {\hat u}^2 \left( \frac{1}{t} -
	\frac{\pi^2}{3 M_C^2} \right ) \left( \frac{1}{\hat s} -
	\frac{\pi^2}{3 M_C^2} \right ) \Biggr ]\;, \nonumber \\
	\overline{\sum} \left| {\cal M} (q\bar q \to gg )\right |^2 &=& g_s^4
	\Biggr \{
	\frac{32}{27} \frac{ {\hat u}^2 + {\hat t}^2}{\hat u \hat t} -
	\frac{8}{3} \frac{ {\hat u}^2 + {\hat t}^2}{\hat s^2} \Biggr \} \;,
      \nonumber \\
	\overline{\sum} \left| {\cal M} (gg \to q\bar q )\right |^2 &=& g_s^4
	\Biggr \{
	\frac{1}{6} \frac{ {\hat u}^2 + {\hat t}^2}{\hat u \hat t} -
	\frac{3}{8} \frac{ {\hat u}^2 + {\hat t}^2}{\hat s^2} \Biggr \} \;,
      \nonumber \\
	\overline{\sum} \left| {\cal M} (q g \to qg )\right |^2 &=& g_s^4
	\Biggr \{ \frac{ {\hat s}^2 + {\hat u}^2}{\hat t^2} - \frac{4}{9}
	\frac{ {\hat s}^2 + {\hat u}^2}{\hat u \hat s} \Biggr \} \;, \nonumber \\
	\overline{\sum} \left| {\cal M} (gg \to gg )\right |^2 &=&
	\frac{9}{4} g_s^4 \Biggr \{ \frac{ {\hat s}^2 + {\hat u}^2 }{ \hat
	t^2} + \frac{ {\hat s}^2 + {\hat t}^2 }{ \hat u^2} + \frac{ {\hat t}^2
	+ {\hat u}^2 }{ \hat s^2} + 3 \Biggr \} \;.\nonumber
\end{eqnarray} 
In the above, if the final state particles are different, the corresponding 
equations need to be symmetrized via $u \leftrightarrow t$ substitution. 
The parton-level differential cross section is given by
$$
	\frac{d\hat \sigma}{d \cos\theta^*} = \frac{1}{32\pi \hat s} \,
	\overline{\sum} |{\cal M}|^2 \;,
$$
where the range of $\cos\theta^*$ is from 0 to 1. This parton-level cross 
section is then convoluted with the parton distribution functions to give 
the total cross section. The above equations are reduced to the SM cross 
sections in the $M_C \to \infty$ limit. The last four equations are the same 
as the SM cross sections, because of the vanishing trilinear gluon vertex
involving two ground-state gluons.

Both CDF \cite{dj-cdf} and D\O~\cite{dj-d0} published data on
dijet production, including invariant mass $M_{jj}$ and angular
distributions. In the fit, we take into account the full correlation
of data points in the data sets, as given by each experiment.  
We normalize the tree-level SM dijet cross section to the low dijet invariant
mass data, $M_{jj}<400$ GeV.

Collider implications of the KK states of the gluon have also been 
considered recently in Ref.~\cite{duane}.

\subsection{Kaluza-Klein States of the Gluon in the $t \bar t$ Production
at the Tevatron}

In Ref. \cite{nandi}, it was shown that the $t\bar t$ production in
Run~2 of the Tevatron can be used to probe the compactification scales
up to $\sim 3$~TeV. In this paper, we consider the sensitivity from the 
existing
Run 1 data by using the tree-level $t\bar t$ production cross section,
including the contribution of the KK states of the gluon in the $q\bar q
\to t\bar t$ channel. (The $gg\to t\bar t$ channel does not have the
triple vertex interaction with two gluons from the SM 3-brane and one
KK state of the gluon in the bulk, as explained in the previous subsection.)

The subprocess cross sections are given by
\begin{eqnarray}
	\frac{d \hat \sigma}{d \cos\theta^*} (q\bar q\to t \bar t)&=&
	\frac{g_s^4 \beta}{72 \pi \hat s} \left( \frac{1}{\hat
	s}-\frac{\pi^2}{3M_C^2} \right )^2 \biggr[ (m_t^2 -\hat t)^2 + (m_t^2
	-\hat u)^2 + 2 \hat s m_t^2 \biggr] \;, \nonumber\\
	\frac{d \hat \sigma}{d \cos\theta^*} (gg\to t \bar t)&=&
	\frac{g_s^4 \beta}{768 \pi \hat s} \biggr \{
	\frac{4}{(\hat t- m_t^2)^2} ( -m_t^4 -3m_t^2 \hat t -m_t^2\hat u +\hat
	u \hat t ) \nonumber \\ 
	&+& \frac{4}{(\hat u- m_t^2)^2} ( -m_t^4
	-3m_t^2 \hat u -m_t^2 \hat t + \hat u \hat t )  \nonumber \\ 
	&+& \frac{m_t^2}{(\hat u-m_t^2)(\hat t -m_t^2)}( 2m_t^2 + \hat t + \hat u )
	+18 \frac{1}{\hat s^2} ( m_t^4 - m_t^2 (\hat t+ \hat u) + \hat t \hat u )
	 \nonumber \\ 
	&+& \frac{9}{\hat t -m_t^2} \frac{1}{\hat s} ( m_t^4 - 2 m_t^2 \hat t +
	\hat u \hat t ) + \frac{9}{\hat u -m_t^2} \frac{1}{\hat
	s} ( m_t^4 - 2 m_t^2 \hat u + \hat u \hat t ) \Biggr \},\nonumber
\end{eqnarray} 
where $\beta = \sqrt{1 - 4m_t^2/\hat s}$ and $\hat s, \hat t, \hat u$ are 
Mandelstam variables. The above cross section is reduced to the SM top 
pair production cross section in the $M_C \to \infty$ limit.

The latest theoretical calculations of the $t\bar t$ cross section,
including higher-order contributions, at $\sqrt{s}=1.8$ TeV correspond
to 4.7 -- 5.5 pb \cite{top-x}.  The present data on the $t\bar t$
cross sections are \cite{tt-data}
\begin{eqnarray}
	\sigma_{t\bar t}\;({\rm CDF}) &=& 6.5 \err{1.7}{1.4} \;\; {\rm pb;}
	\nonumber \\
	\sigma_{t\bar t}\;({\rm \mbox{D\O}})&=& 5.9 \pm{1.7}\;\; {\rm pb,}
	\nonumber
\end{eqnarray} and the top-quark mass measurements are
\begin{eqnarray} 
	m_{t}\;({\rm CDF}) &=& 176.1 \pm 6.6 \;\; {\rm GeV;} \nonumber \\ 
	m_{t}\;({\rm \mbox{D\O}}) &=& 172.1 \pm{7.1}\;\; {\rm GeV} \nonumber \;.
\end{eqnarray} In our analysis, we normalize the tree-level SM cross
section to the mean of the latest theoretical predictions (5.1 pb),
and use this normalization coefficient to predict the cross section in
presence of the KK states of the gluon (similar to Eq. (\ref{sign})).

The effects of KK states of the $W$ boson on single top production 
were recently considered in Ref.~\cite{datta}.

\section{Constraints from High Energy Experiments}
\label{sec:results}

In the previous section, we have described the data sets from various high
energy experiments used in our analysis.
Based on the above individual and combined data sets, we perform a fit to
the sum of the SM prediction and the contribution of the KK states of
gauge bosons, normalizing our tree-level cross section to the best
available higher-order calculations, as explained above. As seen from
Eq. (\ref{mab}), the effects of the KK states always enter the
equations in the form $\pi^2/(3M_C^2)$. Therefore, we parameterize
these effects with a single fit parameter $\eta$:
$$
	\eta = \frac{\pi^2}{3 M_C^2} \;.
$$
In most cases, the differential cross sections in presence of the KK states 
of gauge bosons are bilinear in $\eta$.

The best-fit values of $\eta$ for each individual data set and their
combinations are shown in Table \ref{table1}. In all cases, the preferred
values from the fit are consistent with zero, and therefore we proceed 
with setting limits on $\eta$. The one-sided 95\% C.L. upper limit on 
$\eta$ is defined as: 
\begin{equation}
	0.95 = \frac{\int_0^{\eta_{95}} d \eta \; P(\eta) }
	{\int_0^\infty d \eta \; P(\eta) } \;,\label{maxL}
\end{equation}
where $P(\eta)$ is the fit likelihood function given by
$P(\eta)= \exp( -(\chi^2(\eta) - \chi^2_{\rm min})/2 )$.
The corresponding 
upper 95\% C.L. limits on $\eta$ and lower 95\% C.L. limits on $M_C$ are 
also shown in Table \ref{table1}.

\section{Sensitivity in Run~2 of the Tevatron and at the LHC}
\label{sec:sensitivity}

At the Tevatron, the best channel to probe the KK states of photon or
$Z$ boson is Drell-Yan production. Since the typical $\sqrt{\hat
s}$ in Run~2 is well below the limit obtained in the previous section,
the approximation $M_C^2 \gg \hat s, |\hat t|, |\hat u|$ is still
valid.  Therefore, we can use the reduced amplitudes of
Eq. (\ref{mab}). This approximation also holds well for the LHC, which
was tested by a direct comparison of the approximate cross section
given by Eq. (\ref{mab}) and exact sum over the KK resonances, for
values of $M_C$ $\sim 10$~TeV.

In Ref. \cite{greg}, we showed that using the double differential
distribution $d^2\sigma/ M_{\ell\ell} d \cos\theta$ can increase the
sensitivity to the KK states of the graviton compared to the use of
single-differential distributions.  Similarly, we expect this to be 
the case for the KK states of the photon and the $Z$ boson.  
The double differential 
cross section for Drell-Yan production, including the interactions of 
the KK states of the $\gamma$ and $Z$, is given by
\begin{eqnarray}
	\frac{d^3\sigma}{dM_{\ell\ell} dy d\cos\theta^*} &=& K \sum_q
	\frac{M_{\ell\ell}^3}{192 \pi s} f_q(x_1) f_{\bar q}(x_2)
	\, \biggr[ (1+\cos\theta^*)^2 \left( |M_{LL}^{eq}(\hat s)|^2
	+|M_{RR}^{eq}(\hat s)|^2 \right) \nonumber \\ 
	&&+ (1-\cos\theta^*)^2
	\left( |M_{LR}^{eq}(\hat s)|^2 +|M_{RL}^{eq}(\hat s)|^2 \right) 
	\biggr ] \;,\nonumber
\end{eqnarray} 
where $M_{\alpha\beta}^{eq}$'s are given by Eq. (\ref{mab}), $\theta^*$ 
is the scattering angle in the rest frame of the initial partons, 
$\hat s = M_{\ell\ell}^2$, $dx_1 dx_2 = (2 M_{\ell\ell}/s) dM_{\ell\ell} dy$,
and $x_{1,2} = M_{\ell\ell} e^{\pm y}/\sqrt{s}$.

We follow the prescription of Ref. \cite{greg} and use the Bayesian
approach, which correctly takes into account both the statistical and
systematic uncertainties, in the estimation of the sensitivity to 
$\eta \equiv \pi^2/(3M_C^2)$.
\footnote{Note that the maximum likelihood method, as
given by Eq. (\protect\ref{maxL}), artificially yields ~10\% higher
sensitivity to $M_C$, as it does not properly treat the cases when the
likelihood maximum is found in the unphysical region $\eta<0$.}
Due to the 
high statistics in Run 2 and particularly at the LHC, the overall
systematics becomes dominated by the systematics on the $\hat s$-dependence
of the $K$-factor from the NLO corrections. (Systematic 
uncertainties on the integrated luminosity and efficiencies are not as
important as before, because they get canceled out when normalizing 
the tree level SM cross 
section to the $Z$-peak region in the data.) The uncertainty on the 
$K$-factor from the NLO calculations for Drell-Yan 
production~\cite{vanneerven} 
is currently known to a 3\% level, so we use this as the correlated 
systematics in our calculations on $M_C$. For the LHC we quote the limits 
for the same nominal 3\% uncertainty and also show how the sensitivity 
improves if the uncertainty on the $K$-factor shape is reduced to a 1\% 
level. It shows the importance of higher-order calculations of the 
Drell-Yan cross section, which we hope will become available by the time 
the LHC turns on.
\footnote{The electroweak radiative corrections have recently been 
computed in Ref. \protect\cite{baur}.}

In the simulation, we use a dilepton efficiency of 90\%, a rapidity
coverage of $|\eta| < 2.0$, and typical energy resolutions of the
Tevatron or LHC experiments. The simulation is done for a single
collider experiment in the combination of the dielectron and dimuon
channels.

As expected, the fit to double-differential cross sections yields a $\sim
10\%$ better sensitivity to $M_C$ than just using one-dimensional
differential cross sections.  We illustrate this by calculating the
sensitivity to $M_C$ in Run 1, which is slightly higher than the
result obtained from the fit to the invariant mass spectrum
from CDF and D\O.

The sensitivity, at the 95\% C.L., to $M_C$ in Run 1 (120 pb$^{-1}$), 
Run 2a (2 fb$^{-1}$), Run 2b (15 fb$^{-1}$), and at the LHC (100 fb$^{-1}$)
is given in Table~\ref{table-ll}. While the Run 2 sensitivity is somewhat 
inferior to the current indirect limits from precision electroweak data, 
LHC would offer a significantly higher sensitivity to $M_C$, well above 
10 TeV.

When this work is completed, we learned of a preliminary study on a similar 
topic for the LHC~\cite{LesHouches}, which yielded a somewhat lower
sensitivity. Very recently, a complementary paper~\cite{McMullen} on the 
effects of 
KK excitations of gauge bosons at high-energy $e^+e^-$ colliders 
has appeared in LANL archives.

\section*{\bf Acknowledgments} We would like to thank Ignatios
Antoniadis, Keith Dienes, JoAnne Hewett, Steve Mrenna, Giacomo
Polesello, and Tom Rizzo for useful discussions. This research was
partially supported by the U.S.~Department of Energy under Grants
No. DE-FG02-91ER40688 and by A.P.~Sloan Foundation, and by the 
National Center for Theoretical Science under a grant from the 
National Science Council of Taiwan R.O.C.

\section*{Appendix}

Tables~\ref{tablefirst} to \ref{tablelast} are the data sets that we used
in our analysis.

\begin{table}[!]
\caption{Best-fit values of $\eta=\pi^2/(3M_C^2)$ 
and the 95\% C.L. upper limits on $\eta$ for individual data set 
and combinations. Corresponding 95\% C.L. lower limits on $M_C$ are 
also shown. \label{table1} }
\medskip 
\begin{ruledtabular}
\begin{tabular}{lccc}
      & $\eta$ (TeV$^{-2}$)  & $\eta_{95}$ (TeV$^{-2}$) &  $M_C^{95}$ (TeV) \\
\hline
\hline
LEP~2:                       & & & \\
{} hadronic cross section, ang. dist., $R_{b,c}$
   & $-0.33 \err{0.13}{0.13}$ & 0.12  & 5.3 \\
{} $\mu,\tau$ cross section \& ang. dist.
     & $0.09\err{0.18}{0.18}$ & 0.42 & 2.8 \\
{} $ee$ cross section \& ang. dist.
     & $-0.62\err{0.20}{0.20}$ & 0.16 & 4.5\\
{} LEP combined          & $-0.28\err{0.092}{0.092}$ & 0.076 & 6.6 \\
\hline
HERA:     & & & \\
{} NC     & $-2.74\err{1.49}{1.51}$ & 1.59 & 1.4 \\
{} CC     & $-0.057\err{1.28}{1.31}$ & 2.45 & 1.2 \\
{} HERA combined & $-1.23\err{0.98}{0.99}$ & 1.25  & 1.6 \\
\hline
TEVATRON:              & & & \\
{} Drell-yan           & $-0.87\err{1.12}{1.03}$ & 1.96 & 1.3 \\       
{} Tevatron dijet      & $0.46\err{0.37}{0.58}$ & 1.0  & 1.8 \\
{} Tevatron top production & $-0.53\err{0.51}{0.49}$ & 9.2 & 0.60 \\
{} Tevatron combined   & $-0.38\err{0.52}{0.48}$ & 0.65 & 2.3 \\
\hline
\hline
All combined & $-0.29\err{0.090}{0.090}$ & 0.071 & 6.8 \\
\end{tabular}
\end{ruledtabular}
\end{table}

\begin{table}[!]
\caption{
Sensitivity to the parameter $\eta=\pi^2/3M_C^2$ in Run 1, 
Run 2 of the Tevatron 
and at the LHC, using the dilepton channel. The corresponding 
95\% C.L. lower limits on $M_C$ are also shown.
\label{table-ll} }
\medskip
\begin{ruledtabular}
\begin{tabular}{ccc}
     & $\eta_{95}$ (TeV$^{-2}$) & 95\% C.L. lower limit on $M_C$ (TeV) \\
\hline
\hline
\underline{Run 1 (120 pb$^{-1}$)}& & \\ 
{}Dilepton & 1.62 & 1.4  \\
\hline
\underline{Run 2a (2 fb$^{-1}$)} & &  \\
{}Dilepton & 0.40 & 2.9 \\
\hline
\underline{Run 2b (15 fb$^{-1}$)} & & \\
{}Dilepton & 0.19 & 4.2 \\
\hline
\underline{LHC (14 TeV, 100 fb$^{-1}$, 3\% systematics)} & & \\
{}Dilepton & $1.81\times10^{-2}$ & 13.5 \\
\hline
\underline{LHC (14 TeV, 100 fb$^{-1}$, 1\% systematics)} & & \\
{}Dilepton & $1.37\times10^{-2}$ & 15.5 \\
\end{tabular}
\end{ruledtabular}
\end{table}


\begin{table}[!] 
\caption{
ZEUS: differential cross-section $d\sigma/dQ^2$ of the $e^{+} p
 \rightarrow e^{+} X$
production. The following quantities are given for each bin: the $Q^2$
range, the measured Born-level cross-section, and the SM prediction
for the Born-level cross section.}
\label{tablefirst}
\medskip
{\footnotesize
\begin{ruledtabular}
\begin{tabular}{rrr}
$Q^2$ range & \multicolumn{2}{c}{$d\sigma/dQ^2$ (pb/GeV$^2$)}  \\ 
(GeV$^{2}$) & Measured & SM \\ 
\hline
$400.0-475.7$ &   $ 2.753 \pm  0.035 \;^{+  0.066}_{- 0.051}$ & $2.673$\\
$475.7-565.7$ &   $ 1.753 \pm  0.024 \;^{+  0.047}_{- 0.039}$ & $ 1.775$ \\
$565.7-672.7$ &   $ 1.187 \pm  0.018 \;^{+  0.022}_{- 0.023}$ & $ 1.149$ \\
$672.7-800.0$ &   $(7.71 \pm  0.13 \;^{+  0.14}_{- 0.36}) \cdot 10^{-1}$ & 
$ 7.65 \cdot 10^{-1}$ \\
$800.0-951.4$ &   $(4.79 \pm  0.09 \;^{+  0.10}_{- 0.21})\cdot 10^{-1}$ & 
$ 4.93 \cdot 10^{-1}$ \\
$951.4-1131.4$ &   $(3.21 \pm  0.07 \;^{+  0.06}_{- 0.06})\cdot 10^{-1}$ & 
$ 3.13 \cdot 10^{-1}$ \\
$1131.4-1345.4$ &   $(2.01 \pm  0.05 \;^{+  0.04}_{- 0.03})\cdot 10^{-1}$ & 
$ 2.04 \cdot 10^{-1}$ \\
$1345.4-1600.0$ &   $(1.27 \pm  0.03 \;^{+  0.03}_{- 0.02})\cdot 10^{-1}$ & 
$ 1.28 \cdot 10^{-1}$ \\
$1600.0-1902.7$ &   $(8.49 \pm  0.26 \;^{+  0.17}_{- 0.30})\cdot 10^{-2}$ & 
$ 8.26 \cdot 10^{-2}$ \\
$1902.7-2262.8$ &  $(4.97 \pm  0.18 \;^{+  0.11}_{- 0.16})\cdot 10^{-2}$ & 
$ 5.01 \cdot 10^{-2}$ \\
$2262.8-2690.9$ &  $(3.05 \pm  0.13 \;^{+  0.06}_{- 0.14})\cdot 10^{-2}$ & 
$ 3.13 \cdot 10^{-2}$ \\
$ 2690.9-3200.0$ &   $(1.99 \pm  0.10 \;^{+  0.07}_{- 0.09})\cdot 10^{-2}$ & 
$ 2.09 \cdot 10^{-2}$ \\
$ 3200.0-4525.5$ &   $(9.00 \pm  0.39 \;^{+  0.20}_{- 0.24})\cdot 10^{-3}$ & 
$ 9.77 \cdot 10^{-3}$ \\
$ 4525.5-6400.0$ &   $(3.30 \pm  0.19 \;^{+  0.17}_{- 0.10})\cdot 10^{-3}$ & 
$ 3.49 \cdot 10^{-3}$ \\
$ 6400.0-9050.0$ &   $(1.32 \pm  0.10 \;^{+  0.02}_{- 0.07})\cdot 10^{-3}$ & 
$ 1.20 \cdot 10^{-3}$ \\
$ 9050.0-12800.0$ &   $(3.69 \,^{+  0.53}_{- 0.47} 
\,^{+  0.08}_{- 0.11}) \cdot 10^{-4}$ & $ 3.64 \cdot 10^{-4}$  \\
$12800.0-18102.0$ &   $(8.9 \,^{+  2.5}_{- 2.0} \,^{+  0.7}_{- 0.6})
\cdot 10^{-5}$ &  $ 10.0 \cdot 10^{-5}$  \\
$18102.0-25600.0$ &   $(2.4 \,^{+  1.2}_{- 0.8} \,^{+  0.4}_{- 0.1})
\cdot 10^{-5}$ &  $ 2.2 \cdot 10^{-5}$  \\
$25600.0-36203.0$ & $< 6.0\cdot 10^{-6}$  & $ 3.7 \cdot 10^{-6}$ \\
$36203.0-51200.0$ &  $(2.6 \,^{+  3.5}_{- 1.7} \,^{+  0.7}_{- 0.2})
\cdot 10^{-6}$ &  $ 0.4 \cdot 10^{-6}$  \\
\end{tabular}
\end{ruledtabular}
}
\end{table}


\begin{table}[h]
\caption
{ZEUS: differential cross section $d\sigma/dQ^2$ of the $e^{+} p
\rightarrow \bar{\nu}_e X$ production. The following quantities are
given for each bin: the $Q^2$ range; the measured Born-level cross
section $d\sigma/dQ^2$, and the SM prediction for the Born-level
cross section.}
\medskip
\begin{ruledtabular}
\begin{tabular}{rrr}
$Q^2$ range &  \multicolumn{2}{c}{$d\sigma/dQ^2$(pb/GeV$^2$)}\\ 
(GeV$^2$) &  measured & SM\\ 
\hline
$200 - 400$ & $(2.94 \pm0.28 \,^{+ 0.35}_{- 0.34})\cdot10^{-2}$ & 
$2.80\cdot10^{-2}$\\
$400 - 711$ &   $(1.82 \pm0.14 \pm 0.08) \cdot10^{-2}$ & $1.87\cdot10^{-2}$\\
$711 - 1265$ &  $(1.29 \pm0.08 \pm 0.03) \cdot10^{-2}$ & $1.15\cdot10^{-2}$\\
$1265 - 2249$ &  $(5.62 \pm0.40 \pm 0.08) \cdot10^{-3}$ & $6.07\cdot10^{-3}$\\
$2249 - 4000$ &  $(2.62 \pm0.20 \,^{+ 0.04}_{- 0.09}) \cdot10^{-3}$& 
$2.61\cdot10^{-3}$\\
$4000 - 7113$ & $(7.91 \,^{+ 0.93}_{- 0.83} \,^{+ 0.38}_{- 0.31}) \cdot10^{-4}$&
 $8.29\cdot10^{-4}$\\
$7113 - 12649$ &  $(2.00 \,^{+ 0.35}_{- 0.30}\pm 0.17) \cdot10^{-4}$& 
$1.65\cdot10^{-4}$\\
$12649 - 22494$ &$(2.61 \,^{+ 0.95}_{- 0.72} \,^{+ 0.45}_{- 0.38})\cdot10^{-5}$&
 $1.71\cdot10^{-5}$\\
$22494 - 60000$ & $(5.9 \,^{+ 14.}_{- 4.9} \,^{+ 1.8}_{- 1.5})\cdot10^{-7}$& 
$6.24\cdot10^{-7}$
\end{tabular}
\end{ruledtabular}
\end{table}

\begingroup
\squeezetable
\begin{table}[ht]
\caption{H1: reduced NC cross-section $\tilde{\sigma}_{\rm NC}(x,Q^2)$
in the $e^+ p$ collisions obtained by dividing $d^2\sigma_{\rm NC}/dx
dQ^2$ by the kinematic factor $x Q^4/(Y_+ 2\pi \alpha^2)$, with its
statistical ($\delta_{\rm stat}$), systematic ($\delta_{\rm sys}$),
and combined ($\delta_{\rm tot}$) uncertainties. The additional
normalization uncertainty, not included in the systematic error, is
$1.5\%$. The table continues on the next two pages.}
\tiny
\begin{ruledtabular}
\begin{tabular}{ccccrrr}
      $Q^2$  &$x$  & $y$ & $\tilde{\sigma}_{\rm NC}$ & 
      $\delta_{\rm stat}$ & $\delta_{\rm sys}$ &
      $\delta_{\rm tot}$        \\[-1mm]
       $(\rm GeV^2)$ & & & &
       $(\%)$ &
       $(\%)$ & $(\%)$  \\
 \hline
$   150$&$  0.003$&$  0.518$&$  1.240$&$  1.8$&$  5.2$&$  5.5$ \\
$   150$&$  0.005$&$  0.331$&$  1.100$&$  1.8$&$  3.3$&$  3.8$ \\
$   150$&$  0.008$&$  0.207$&$  0.920$&$  2.9$&$  8.9$&$  9.3$ \\
 \hline                                                            
$   200$&$  0.005$&$  0.442$&$  1.102$&$  1.8$&$  5.0$&$  5.3$ \\
$   200$&$  0.008$&$  0.276$&$  0.915$&$  1.9$&$  3.5$&$  4.0$ \\
$   200$&$  0.013$&$  0.170$&$  0.765$&$  2.2$&$  3.7$&$  4.3$ \\
\end{tabular}
\end{ruledtabular}
\end{table}
\endgroup

\begingroup
\squeezetable
\begin{table*}[ht]
\tiny
\begin{ruledtabular}
    \begin{tabular}{cccrrrr}
      $Q^2$  &$x$  & $y$ & $\tilde{\sigma}_{\rm NC}$ & 
      $\delta_{\rm stat}$ & $\delta_{\rm sys}$ &
      $\delta_{\rm tot}$       \\[-1mm]
       $(\rm GeV^2)$ & & & &
       $(\%)$ &
       $(\%)$ & $(\%)$ \\
 \hline
$   500$&$  0.008$&$  0.690$&$  1.026$&$  4.2$&$  5.1$&$  6.6$ \\
$   500$&$  0.013$&$  0.425$&$  0.906$&$  3.3$&$  5.2$&$  6.2$ \\
$   500$&$  0.020$&$  0.276$&$  0.792$&$  3.3$&$  3.9$&$  5.2$ \\
$   500$&$  0.032$&$  0.173$&$  0.654$&$  3.3$&$  4.0$&$  5.2$ \\
$   500$&$  0.050$&$  0.110$&$  0.508$&$  3.5$&$  4.1$&$  5.4$ \\
$   500$&$  0.080$&$  0.069$&$  0.445$&$  3.6$&$  3.7$&$  5.2$ \\
\end{tabular}
\end{ruledtabular}
\end{table*}
\endgroup

\begingroup
\squeezetable
\begin{table*}[ht]
\tiny
\begin{ruledtabular}
    \begin{tabular}{cccrrrr}
      $Q^2$  &$x$  & $y$ & $\tilde{\sigma}_{\rm NC}$ & 
      $\delta_{\rm stat}$ & $\delta_{\rm sys}$ &
      $\delta_{\rm tot}$        \\[-1mm]
       $(\rm GeV^2)$ & & & &
       $(\%)$ &
       $(\%)$ & $(\%)$ \\
 \hline                                                          
$  1500$&$  0.020$&$  0.828$&$  0.789$&$  9.2$&$  5.0$&$ 10.5$ \\
$  1500$&$  0.032$&$  0.518$&$  0.581$&$  8.1$&$  4.3$&$  9.2$ \\
$  1500$&$  0.050$&$  0.331$&$  0.486$&$  7.2$&$  3.8$&$  8.1$ \\
$  1500$&$  0.080$&$  0.207$&$  0.457$&$  6.8$&$  3.7$&$  7.8$ \\
$  1500$&$  0.130$&$  0.127$&$  0.376$&$  8.0$&$  3.9$&$  8.9$ \\
$  1500$&$  0.180$&$  0.092$&$  0.345$&$  8.6$&$  4.2$&$  9.6$ \\
\end{tabular}
\end{ruledtabular}
\end{table*}
\endgroup


\begingroup
\squeezetable
\begin{table}[ht]
\caption{H1: double differential CC cross-section $d^2\sigma_{\rm
CC}/dx dQ^2$ with its statistical ($\delta_{\rm stat}$), systematic
($\delta_{\rm sys}$), and combined ($\delta_{\rm tot}$) uncertainties
in the $e^+ p$ collisions. The additional normalization uncertainty,
not included in the systematic error, is 1.5\%.}
\begin{ruledtabular}
    \begin{tabular}{cccrrrr}
      $Q^2$  &$x$  & $y$ & $d^2\sigma_{\rm CC}/dxdQ^2$ &  
      $\delta_{\rm stat}$ & $\delta_{\rm sys}$ & $\delta_{\rm tot}$ \\
      \footnotesize $(\rm GeV^2)$ & &
      & \footnotesize$(\rm pb/GeV^2)$ &\footnotesize $(\%)$ &
      \footnotesize $(\%)$ &\footnotesize $(\%)$ \\
\hline
$   300$&$  0.013$&$  0.255$&\phantom{c} $ 0.637\cdot 10^{0}$ & $ 27.4$ &
  $ 16.0$&$ 31.8$ \\
$   300$&$  0.032$&$  0.104$&\phantom{c} $ 0.124\cdot 10^{0}$ & $ 28.1$ &
  $ 10.3$&$ 30.0$\\
$   300$&$  0.080$&$  0.041$&\phantom{c} $ 0.532\cdot 10^{-1}$& $ 23.8$&
 $  7.5$&$ 25.5$\\
 \hline
$   500$&$  0.013$&$  0.425$&\phantom{c} $ 0.468\cdot 10^{0}$&$ 25.1$ &
 $ 15.7$&$ 29.7$ \\
$   500$&$  0.032$&$  0.173$&\phantom{c} $ 0.177\cdot 10^{0}$&$ 17.0$&
 $  8.7$&$ 19.2$ \\
$   500$&$  0.080$&$  0.069$&\phantom{c} $ 0.546\cdot 10^{-1}$& $ 17.0$&
 $  6.5$&$ 18.9$ \\
\end{tabular}
\end{ruledtabular}
\end{table}
\endgroup

\begingroup
\squeezetable
\begin{table}[htb]
\caption
 {H1: reduced NC cross section $\tilde{\sigma}_{\rm NC}(x,Q^2)$ with
 its combined $(\delta_{\rm tot})$, statistical $(\delta_{\rm stat})$,
 and systematic $(\delta_{\rm sys}$) uncertainties in the $e^- p$
 collisions.  The additional normalization uncertainty of 1.8\% is not
 included in the errors.  The table continues on the next two pages.}
\tiny
\begin{ruledtabular}
 \begin{tabular}{ccrrrr}
 $Q^2$  &$x$ & $\tilde{\sigma}_{\rm NC}$ &
 $\delta_{\rm tot}$ & $\delta_{\rm stat}$ & $\delta_{\rm sys}$  \\
 $(\rm GeV^2)$ & & &  $(\%)$ &$(\%)$ &$(\%)$ \\
\hline
$  150$&$0.0032$&$1.218$&$  4.7$&$  2.7$&$  3.8$ \\ 
$  150$&$0.0050$&$1.154$&$  4.4$&$  2.8$&$  3.4$ \\
$  150$&$0.0080$&$0.968$&$  9.1$&$  4.1$&$  8.2$ \\ 
\hline
$  200$&$0.0032$&$1.271$&$  6.1$&$  4.1$&$  4.5$ \\
$  200$&$0.0050$&$1.107$&$  4.6$&$  2.8$&$  3.6$ \\
$  200$&$0.0080$&$0.915$&$  4.5$&$  3.0$&$  3.3$ \\
\end{tabular}
\end{ruledtabular}
 \end{table}
\endgroup

\begingroup
	\squeezetable
 \begin{table*}[htb]
\tiny
\begin{ruledtabular}
 \begin{tabular}{ccrrrr}
 $Q^2$  &$x$ & $\tilde{\sigma}_{\rm NC}$ &
 $\delta_{\rm tot}$ & $\delta_{\rm stat}$ & $\delta_{sys}$ \\
 $(\rm GeV^2)$ & & &  $(\%)$ &$(\%)$ &$(\%)$ \\
\hline
$  500$&$0.0080$&$1.044$&$  9.3$&$  7.8$&$  5.1$ \\
$  500$&$0.0130$&$1.003$&$  6.8$&$  5.1$&$  4.5$ \\
$  500$&$0.0200$&$0.765$&$  7.0$&$  5.1$&$  4.8$ \\
$  500$&$0.0320$&$0.604$&$  7.0$&$  5.3$&$  4.5$ \\
$  500$&$0.0500$&$0.517$&$  6.9$&$  5.6$&$  4.0$ \\
$  500$&$0.0800$&$0.392$&$  9.2$&$  6.4$&$  6.5$ \\
\end{tabular}
\end{ruledtabular}
 \end{table*}
\endgroup

\begingroup
	\squeezetable
 \begin{table*}[htb]
\tiny
\begin{ruledtabular}
 \begin{tabular}{ccrrrr}
 $Q^2$  &$x$ & $\tilde{\sigma}_{\rm NC}$ &
 $\delta_{\rm tot}$ & $\delta_{\rm stat}$ & $\delta_{\rm sys}$ \\
 $(\rm GeV^2)$ & & &  $(\%)$ &$(\%)$ &$(\%)$ \\
\hline
$ 1500$&$0.0200$&$0.860$&$ 13.5$&$ 12.4$&$  5.5$ \\
$ 1500$&$0.0320$&$0.704$&$ 11.4$&$ 10.4$&$  4.7$ \\
$ 1500$&$0.0500$&$0.515$&$ 12.2$&$ 11.7$&$  3.6$ \\
$ 1500$&$0.0800$&$0.512$&$ 11.7$&$ 11.0$&$  4.0$ \\
$ 1500$&$0.1300$&$0.390$&$ 14.8$&$ 13.9$&$  5.0$ \\
$ 1500$&$0.1800$&$0.260$&$ 19.1$&$ 18.6$&$  4.3$ \\
\end{tabular}
\end{ruledtabular}
 \end{table*}
\endgroup

\begingroup
	\squeezetable
\begin{table}[htb]
\caption{H1: double differential CC cross section $d^2\sigma_{\rm
CC}/dx dQ^2$ with its overall ($\delta_{\rm tot}$), statistical
($\delta_{\rm stat}$),
  and systematic uncertainties ($\delta_{\rm sys}$) in the $e^- p$
  collisions.  The additional normalization uncertainty of 1.8\% is
  not included in the errors.
}
\scriptsize
\begin{ruledtabular}
    \begin{tabular}{ccrrrr}
$Q^2$  & $x$ & 
$d^2 \sigma_{\rm CC} /dx dQ^2$ & 
      $\delta_{\rm tot}$ & $\delta_{\rm stat}$ & $\delta_{\rm sys}$ \\
      $(\rm GeV^2)$ & & $(\rm{pb/GeV^2})$ & 
       $(\%)$ & $(\%)$ &  $(\%)$ \\
 \hline
$  300$ & $0.013$ & \phantom{C} $0.458\cdot 10^{ 0}$ & $57.6$ & $55.4$ &
$ 15.7$ \\
$  300$ & $0.032$ & \phantom{C} $0.399\cdot 10^{ 0}$ & $27.3$ & $24.5$ &
$ 12.0$ \\
$  300$ & $0.080$ & \phantom{C} $0.690\cdot 10^{-1}$ & $42.3$ & $40.7$ &
$ 11.6$ \\
\hline
$  500$ & $0.013$ & \phantom{C} $0.433\cdot 10^{ 0}$ & $39.9$ & $37.6$ &
$ 13.3$ \\
$  500$ & $0.032$ & \phantom{C} $0.285\cdot 10^{ 0}$ & $21.0$ & $19.6$ &
$ 7.8$ \\
$  500$ & $0.080$ & \phantom{C} $0.790\cdot 10^{-1}$ & $22.4$ & $21.8$ &
$ 5.1$ \\
\end{tabular}
\end{ruledtabular}
\end{table}
\endgroup

\begingroup
\squeezetable
\begin{table}[th]
\caption{CDF: Drell-Yan data in the muon and electron channels.}
\scriptsize
\begin{ruledtabular}
\begin{tabular}{cc}
\multicolumn{2}{c}{Muon channel}\\
Mass bin &  $d^2 \sigma/ dM dy |_{|y|<1}$  (pb/GeV) \\
\hline
$40-50$  & $0.367 \pm 0.057$ \\
$50-60$  & $0.129 \pm 0.030$ \\
$60-70$  & $0.107 \pm 0.019$ \\
$70-78$  & $0.124 \pm 0.019$ \\
$78-86$  & $0.360 \pm 0.037$ \\
$86-90$  & $2.36 \pm 0.21$ \\
$90-94$  & $12.38 \pm 1.08$ \\
$94-102$  & $0.550 \pm 0.052$ \\
$102-110$  & $0.161 \pm 0.025$ \\
$110-120$  & $0.069 \pm 0.016$ \\
$120-150$  & $0.024 \pm 0.005$ \\
$150-200$  & $0.0047 \pm 0.0016$ \\
$200-250$  & $0.0021 \pm 0.0011$ \\
$250-300$  & $0.00107 \pm 0.00076$ \\
$300-400$  & $0.00024 \pm 0.00024$ \\
$400-500$  & $0.0 \pm 0.00019 $
\end{tabular}
\medskip
\begin{tabular}{ccc}
\multicolumn{3}{c}{Electron channel}\\
Mass bin & $M_{\rm mean}$  (GeV) & $d \sigma/ dM$  (pb/GeV) \\
\hline
\hline
$40-50$ & $44.5$ & $(2.300 \pm 0.448 \pm 0.126 )\times 10^0$ \\
$50-60$ & $54.6$ & $(8.038 \pm 0.978 \pm 0.478 )\times 10^{-1}$ \\
$60-70$ & $63.8$ & $(6.005 \pm 0.547 \pm 0.293 )\times 10^{-1}$ \\
$70-78$ & $74.7$ & $(5.829 \pm 0.439 \pm 0.148 )\times 10^{-1}$ \\
$78-86$ & $82.9$  & $(1.638 \pm 0.061 \pm 0.011 )\times 10^0$ \\
$86-88$ & $87.1$ & $(5.895 \pm 0.222 \pm 0.027 )\times 10^0$ \\
$88-90$ & $89.2$ & $(1.828 \pm 0.051 \pm 0.009 )\times 10^1$ \\
$90-92$ & $90.6$  &  $(5.485 \pm 0.135 \pm 0.033 )\times 10^1$ \\
$92-94$ & $92.8$  & $(2.275 \pm 0.065 \pm 0.015 )\times 10^1$ \\
$94-100$ & $96.2$  & $(3.796 \pm 0.117 \pm 0.019 )\times 10^0$ \\
$100-105$  & $102.2$ & $(9.120 \pm 0.733 \pm 0.126 )\times 10^{-1}$ \\
$105-120$  & $111.2$ & $(2.631 \pm 0.237 \pm 0.044 )\times 10^{-1}$ \\
$120-140$  & $128.8$ & $(6.554 \pm 1.017 \pm 0.191 )\times 10^{-2}$ \\
$140-200$  & $164.2$ & $(2.083 \pm 0.320 \pm 0.051 )\times 10^{-2}$ \\
$200-300$  & $240.6$ & $(2.599 \pm 0.847 \pm 0.039 )\times 10^{-3}$ \\
$300-400$  & $342.2$ & $(8.080 \pm 4.677 \pm 0.127 )\times 10^{-4}$ \\
$400-600$  & $478.8$ & $(1.433 \pm 1.433 \pm 0.037 )\times 10^{-4}$ \\
$>600$     & $725.6$ & $(0.000 \pm 0.964 \pm 0.000 )\times 10^{-4}$ \\
\end{tabular}
\end{ruledtabular}
\end{table}
\endgroup

\begin{table}[th]
\caption{D\O: Drell-Yan data in the electron channel.}
\begin{ruledtabular}
\begin{tabular}{cc}
\multicolumn{2}{c}{Electron channel}\\
$M_{\ell\ell}$ range (GeV) & $\sigma$ (pb)  \\
\hline
120--160 & 1.93 \err{0.43}{0.44} \\ 
160--200 & 0.49 \err{0.16}{0.18} \\ 
200--240 & 0.28 \err{0.09}{0.10} \\ 
240--290 & 0.066 \err{0.052}{0.058} \\  
290--340 & 0.033 \err{0.032}{0.030} \\ 
340--400 & 0.057 \err{0.042}{0.047} \\ 
400--500 & $<0.063 \;(0.039)$ \\
500--600 & $<0.060 \;(0.037)$ \\
600--1000 & $<0.058 \;(0.035)$
\end{tabular}
\end{ruledtabular}
\end{table}

\begingroup
\squeezetable
\begin{table}[ht]
 \caption{Preliminary combined LEP results on the $e^- e^+ \to f \bar
f$ production at $\sqrt{s}=130-202$ GeV. The Standard Model
predictions are from ZFITTER~\protect\cite{ZFITTER} v6.10.
}
\tiny
\begin{ruledtabular}	
 \begin{tabular}{ccrr}
$\sqrt{s}$ (GeV) & Quantity & Value   & SM  \\
\hline
  130 & $\sigma(q\overline{q})$ \hfill [pb] & $81.938\pm   2.220$ & 82.803 \\
      & $\sigma(\mu^{+}\mu^{-})$ [pb]       & $ 8.592\pm   0.682$ &  8.439 \\
      & $\sigma(\tau^{+}\tau^{-})$ [pb]     & $ 9.082\pm   0.931$ &  8.435 \\
      & $\mathrm{A_{fb}}(\mu^{+}\mu^{-})$   & $ 0.692\pm   0.060$ &  0.705 \\
      & $\mathrm{A_{fb}}(\tau^{+}\tau^{-})$ & $ 0.663\pm   0.076$ &  0.704 \\
\hline						      
  136 & $\sigma(q\overline{q})$ \hfill [pb] & $66.570\pm   1.967$ & 66.596 \\
      & $\sigma(\mu^{+}\mu^{-})$ [pb]       & $ 8.231\pm   0.678$ &  7.281 \\
\end{tabular}
\end{ruledtabular}
\end{table}
\endgroup

\begin{table}[th]
\caption{
Combined LEP~2 Data on the $R_b$ and $R_c$ along with the Standard
 Model predictions.}
\begin{ruledtabular}
\begin{tabular}{lcc}
$\sqrt{s}$ (GeV) & $R_b$                  & $R_c$ \\
\hline
133      & $0.1809 \pm 0.0133 $   & -   \\
         & (0.1853)  & -  \\
167      & $0.1479 \pm 0.0127 $   &  -  \\
         & (0.1708)  & -    \\
183      & $0.1616 \pm 0.0101 $   & $0.270 \pm 0.043$  \\
         & (0.1671)  & (0.250)  \\
189      & $0.1559 \pm 0.0066 $   & $0.241 \pm 0.024$  \\
         & (0.1660)  & (0.252) \\
\hline
192      & $0.1688 \pm 0.0187 $  &  -  \\
         & (0.1655)  &   - \\ 
196      & $0.1577 \pm 0.0109 $   &  -  \\
         & (0.1648)  &   - \\ 
200      & $0.1621 \pm 0.0111 $   &  -  \\
         & (0.1642)  &   - \\ 
202      & $0.1873 \pm 0.0177$    &  -  \\
         & (0.1638)  &   - \\ 
206      & $0.1696 \pm 0.0182$   &  -  \\
         & (0.1633)  &   -  \\ 
\end{tabular}
\end{ruledtabular}
\end{table}

\begingroup
\squeezetable
\begin{table}[ht]
\caption{
The LEP~2 $e^+ e^- \to e^+ e^-$ production cross sections.}
\tiny
\begin{ruledtabular}
\begin{tabular}{ccc}
$\sqrt{s}$ (GeV) & $\sigma_{ee}$ (pb)& $\sigma^{\rm SM}_{ee}$ (pb) \\
\hline
\multicolumn{3}{c}{ALEPH} \\
\hline
130 & $191.3\pm6.2\pm3.5$ & 186.7 \\
136 & $162.2\pm5.6\pm3.5$ & 167.3 \\
161 & $119.7\pm3.7\pm2.3$ & 119.0 \\
172 & $107.8\pm3.5\pm2.1$ & 102.5 \\
183 & $90.9\pm1.4\pm1.7$ & 90.9  \\
189 & $84\pm0.72\pm0.71$ & 87.25  \\
192 & $80.00\pm 1.90 \pm0.87$ & 81.9 \\
196 & $78.74\pm 1.10 \pm 0.76$ & 79.2 \\
200 & $73.95 \pm1.02 \pm 0.70$ & 76.3 \\
202 & $75.42 \pm 1.50 \pm 0.80$ & 74.6 \\
204.9 & $75.33\pm1.20 \pm 0.90$ & 74.36 \\
206.7 & $73.40\pm0.91 \pm 0.70$ & 74.50 \\
\hline
\multicolumn{3}{c}{DELPHI}\\
\hline
130.2 & $42.0\pm4.0\pm0.80$ & 48.7 \\
136.2 & $47.1\pm4.2\pm0.71$ & 44.6 \\
161.3 & $27.1\pm1.8\pm0.41$ & 31.9 \\
172.1 & $30.3\pm1.9\pm0.36$ & 28.0 \\
183 &   $25.6 \pm 0.8$  & 24.7 \\
189 &   $22.6 \pm 0.4$  & 23.1 \\
192 &   $24.03 \pm 1.03$ & 22.29 \\
196 &   $22.43 \pm 0.56$ & 21.36 \\
200 &   $20.56 \pm 0.51$ & 20.52 \\
202 &   $21.30 \pm 0.75$ & 20.08 \\
\hline
\multicolumn{3}{c}{L3}\\
\hline
130.1 & $45.0\pm 2.7 \pm0.2$ & 49.7 \\
136.1 & $43.6 \pm 2.8 \pm 0.2$ & 45.4 \\
161.3 & $31.1\pm1.8 \pm 0.9$ & 32.4 \\
172.3 & $26.7\pm1.8 \pm 0.8$ & 28.3 \\ 
182.7 & $25.6 \pm 0.7 \pm 0.1$ & 25.0 \\
188.7 & $23.5 \pm 0.4 \pm 0.1$ & 23.4 \\
\hline
\multicolumn{3}{c}{OPAL} \\
\hline
130.25 & $615\pm16\pm8$ & 645 \\
136.22 & $580\pm15\pm8$ & 592 \\
161.34 & $434\pm7\pm5$  & 425 \\
172.12 & $365\pm6\pm5$  & 375 \\
183    & $333\pm3\pm4$ & 333  \\
189    & $304.6\pm1.3\pm1.4$ & 311.6 \\
192  & $301.4\pm 3.3 \pm 1.5$ & 299.4 \\
196  & $285.8\pm 2.0 \pm 1.5$ & 287.7 \\
200  & $273.0\pm 1.9 \pm 1.4$ & 276.3 \\
202  & $272.0\pm2.8 \pm 1.4$ & 270.6 
\end{tabular}
\end{ruledtabular}
\end{table}
\endgroup


\begin{table}[ht]
\caption{\label{lep-ang-al}
The $e^+ e^- \to e^+ e^-$ differential cross section $d\sigma/d\cos\theta^*$ 
measured by ALEPH.}
\medskip
\begin{ruledtabular}
\begin{tabular}{ccccccc}
$\cos\theta^*$ range & \multicolumn{2}{c}{130 GeV}  & 
       \multicolumn{2}{c}{136 GeV} & \multicolumn{2}{c}{161 GeV} \\
 & $\sigma$ (pb) & $\sigma_{\rm SM}$ (pb) &
   $\sigma$ (pb) & $\sigma_{\rm SM}$ (pb) &
   $\sigma$ (pb) & $\sigma_{\rm SM}$ (pb)   \\
\hline
\hline
$-0.9 - -0.7$ & $0.19\pm0.34$ & 0.37 & $0.73\pm0.20$ & 0.22 & 
    $0.46\pm0.21$ & 0.37  \\
$-0.7 - -0.5$ & $1.41\pm0.35$ & 0.55 & $1.16\pm0.36$ & 0.62 & 
    $0.88\pm0.21$ & 0.44  \\
$-0.5 - -0.3$ & $1.36\pm0.45$ & 1.09 & $0.54\pm0.35$ & 0.49 & 
    $0.55\pm0.28$ & 0.79  \\
$-0.3 - -0.1$ & $1.23\pm0.48$ & 1.19 & $0.52\pm0.41$ & 0.89 & 
    $0.39\pm0.26$ & 0.62  \\
$-0.1 -  0.1$ & $2.60\pm0.69$ & 2.45 & $1.46\pm0.62$ & 2.09 & 
    $1.24\pm0.40$ & 1.43  \\
$ 0.1 -  0.3$ & $3.78\pm0.83$ & 3.82 & $2.09\pm0.74$ & 2.96 & 
    $2.37\pm0.47$ & 2.07  \\
$ 0.3 -  0.5$ & $8.88\pm1.18$ & 7.36 & $6.68\pm1.08$ & 6.13 & 
    $5.35\pm0.73$ & 4.95  \\
$ 0.5 -  0.7$ & $21.63\pm2.12$ & 22.20 & $16.58\pm1.97$ & 20.50 & 
    $14.38\pm1.27$ & 14.10  \\
$ 0.7 -  0.9$ & $149.61\pm6.22$ & 148.0 & $132.55\pm5.85$ & 133.0 & 
    $93.76\pm3.76$ & 94.20  \\
\hline
 & \multicolumn{2}{c}{172 GeV} & \multicolumn{2}{c}{183 GeV} &
   \multicolumn{2}{c}{}  \\
\hline
\hline
$-0.9 - -0.7$ & $0.32\pm0.19$ & 0.28 & $0.24\pm0.07$ & 0.21 & & \\
$-0.7 - -0.5$ & $0.88\pm0.19$ & 0.34 & $0.29\pm0.07$ & 0.25 & & \\
$-0.5 - -0.3$ & $0.66\pm0.24$ & 0.58 & $0.46\pm0.10$ & 0.51 &  & \\
$-0.3 - -0.1$ & $0.61\pm0.23$ & 0.44 & $0.71\pm0.12$ & 0.64 &  & \\
$-0.1 -  0.1$ & $0.95\pm0.36$ & 1.23 & $0.83\pm0.14$ & 0.90 &  & \\
$ 0.1 -  0.3$ & $1.80\pm0.47$ & 1.93 & $1.42\pm0.20$ & 1.83 &  & \\
$ 0.3 -  0.5$ & $4.92\pm0.71$ & 4.24 & $3.90\pm0.29$ & 3.66 &  & \\
$ 0.5 -  0.7$ & $13.07\pm1.20$ & 12.40 & $12.47\pm0.56$ & 11.10 &  & \\
$ 0.7 -  0.9$ & $84.61\pm3.51$ & 81.10 & $71.90\pm1.86$ & 71.80 &  & \\
\end{tabular}
\end{ruledtabular}
\end{table}

\begin{table}[ht]
\caption{
The $e^+ e^- \to \mu^+ \mu^-$ and $e^+ e^- \to \tau^+ \tau^-$ differential 
cross sections $d\sigma/d\cos\theta^*$ measured by ALEPH.}
\medskip
\begin{ruledtabular}
\begin{tabular}{cccccccc}
$\cos\theta^*$ range & \multicolumn{7}{c}{$d\sigma/d\cos\theta^*$ (pb)} \\
  & 189 GeV & 192 GeV & 196 GeV & 200 GeV & 202 GeV &204.9 GeV& 206.7 GeV \\
\hline
\multicolumn{8}{c}{$\mu^+ \mu^-$} \\
\hline
$-0.95$-$-0.8$&$0.69\pm0.17$&$0.23\pm0.25$ &$0.61\pm0.24$ & $0.64\pm0.23$ &
$0.83\pm0.38$ & $0.42\pm0.20$ & $0.51\pm0.18$ \\
$-0.8$-$-0.6$&$0.31\pm0.10$&$0.17\pm0.18$ &$0.56\pm0.20$ & $0.58\pm0.19$ &
$0.10\pm0.13$ & $0.29\pm0.15$ & $0.43\pm0.14$ \\
$-0.6$-$-0.4$&$0.38\pm0.11$&$1.07\pm0.45$ &$0.51\pm0.19$ & $0.90\pm0.24$ &
$0.86\pm0.33$ & $0.17\pm0.13$ & $0.25\pm0.11$ \\
$-0.4$-$-0.2$&$0.75\pm0.16$&$0.72\pm0.37$ &$0.58\pm0.20$ & $1.02\pm0.26$ &
$0.48\pm0.26$ & $0.56\pm0.31$ & $1.00\pm0.21$ \\
$-0.2$-$0.0$&$1.11\pm0.19$&$0.89\pm0.42$ &$0.88\pm0.25$ & $1.20\pm0.28$ &
$0.73\pm0.32$ & $0.94\pm0.25$ & $0.60\pm0.16$ \\
\end{tabular}
\end{ruledtabular}
\end{table}

\begin{table}[ht]
\caption{\small
Differential cross section $d\sigma/d\cos\theta^*$ for the 
$e^+ e^- \to e^+ e^-, \mu^+\mu^-,
\tau^+\tau^-, q\bar q$ production measured by OPAL.}
\medskip
\tiny
\begin{ruledtabular}
\begin{tabular}{ccccccccc}
$\cos\theta^*$ range & \multicolumn{8}{c}{$d\sigma/d\cos\theta^*$ (pb)} \\
  & 130.12 GeV  & 136.08 GeV & 182.69 GeV & 189 GeV & 192 GeV & 196 GeV &
  200 GeV & 202 GeV\\[6pt]
\hline
 & \multicolumn{8}{c}{$e^+ e^-$} \\
\hline
$-0.9 - -0.7$&$4\err{3}{2}$&$5\err{3}{2}$ &$1.2\err{0.4}{0.3}$ & $1.4\pm0.2$
 & $1.7\err{0.7}{0.5}$ & $1.4\pm0.3$ & $1.6\pm0.3$ & $2.3\err{0.7}{0.6}$\\
$-0.7 - -0.5$&$4\err{3}{2}$&$4\err{3}{2}$ &$2.1\pm{0.4}$ & $2.0\pm0.2$
 & $1.9\err{0.8}{0.6}$ & $1.6\pm0.3$ & $1.9\pm0.3$ & $1.6\err{0.6}{0.5}$ \\
$-0.5 - -0.3$&$6\err{3}{2}$&$8\err{4}{3}$ &$2.3\pm{0.5}$ & $2.4\pm0.3$
  &$1.9\err{0.8}{0.6}$ & $2.6\pm0.4$ & $2.3\pm0.4$ & $3.2\pm0.7$ \\
$-0.3 - -0.1$&$6\err{4}{2}$&$9\err{4}{3}$ &$4.8\pm{0.7}$ & $3.0\pm0.3$
  &$1.5\err{0.7}{0.5}$ & $2.7\pm0.4$ & $2.4\pm0.4$ & $2.5\err{0.7}{0.6}$ \\
$-0.1 -  0.1$&$13\err{5}{4}$&$8\err{4}{3}$ &$6.1\pm{0.7}$ & $4.3\pm0.3$
  &$5.4\pm1.0$ & $3.7\pm0.5$ & $5.2\pm0.6$ & $4.4\pm0.8$ \\
\end{tabular}
\end{ruledtabular}
\end{table}


\begingroup
\squeezetable
\begin{table}[ht]
\caption{
Differential cross section $d\sigma/d\cos\theta^*$ for the 
$e^+ e^- \to \mu^+\mu^-, \; \tau^+\tau^-$ production measured by DELPHI.
For $\sqrt{s}=183$ GeV the $\cos\theta^*$ range for $\mu^+ \mu^-$ channel
is from $-0.94$ to $0.94$.}
\medskip
\tiny
\begin{ruledtabular}
\begin{tabular}{lcccccc}
$\cos\theta^*$ range & \multicolumn{6}{c}{$d\sigma/d\cos\theta^*$ (pb)} \\
  & 183 GeV & 189 GeV & 192 GeV & 196 GeV & 200 GeV & 202 GeV \\
\hline
 & \multicolumn{6}{c}{$\mu^+ \mu^-$} \\
\hline
$-0.97$-$-0.8$ & $.000\pm.178\pm.013$ & $.495\pm.143\pm.008$ & 
      $.000\pm.331\pm.005$ & $.716\pm.253\pm.018$ &
      $.560\pm.211\pm.016$ & $1.077\pm.407\pm.029$ \\
$-0.8$-$-0.6$ & $.514\pm.230\pm.013$ & $.478\pm.128\pm.008$ &
      $.202\pm.202\pm.005$ & $.520\pm.196\pm.013$ &
      $.266\pm.133\pm.007$ & $.585\pm.292\pm.017$ \\
$-0.6$-$-0.4$ & $.989\pm.313\pm.024$ & $.448\pm.120\pm.007$ & 
    $.814\pm.407\pm.020$ & $.614\pm.205\pm.015$ &
    $.885\pm.236\pm.023$ & $.259\pm.183\pm.007$ \\
$-0.4$-$-0.2$& $ .972\pm.307\pm.023$ & $.391\pm.113\pm.006$ &
    $.385\pm.272\pm.009$ & $.208\pm.120\pm.005$ &
    $.699\pm.211\pm.018$ & $.249\pm.176\pm.007$ \\
$-0.2$-$0.0$&$1.298\pm.360\pm.032$ & $1.287\pm.212\pm.021$ &
    $1.068\pm.477\pm.027$ & $.875\pm.253\pm.022$ &
    $1.053\pm.263\pm.029$ & $.392\pm.226\pm.011$ \\
\end{tabular}
\end{ruledtabular}
\end{table}
\endgroup


\begin{table}[ht]
\caption{
Differential cross section $d\sigma/d\cos\theta^*$ for the $e^+ e^-
\to e^+ e^-,\,\mu^+\mu^-, \; \tau^+\tau^-$ production measured by L3.}
\scriptsize
\begin{ruledtabular}
\begin{tabular}{lcc}
$\cos\theta^*$ range & \multicolumn{2}{c}{$d\sigma/d\cos\theta^*$ (pb)} \\
  & 182.7 GeV & 188.7 GeV \\
\hline
  \multicolumn{3}{c}{$e^+ e^-$} \\
\hline
$-0.719 - -0.575$ & $0.404\pm0.157$ & $0.167\pm0.075$ \\
$-0.575 - -0.432$ & $0.306\pm0.115\pm0.013$ & $0.300\pm0.065$ \\
$-0.432 - -0.288$ & $0.532\pm0.127$ & $0.526\pm0.070$ \\
$-0.288 - -0.144$& $ 0.539\pm0.122$ & $0.524\pm0.067$ \\
$-0.144 -  0.0$ & $0.930\pm0.153$ & $0.708\pm0.075$ \\
$ 0.0 -  0.144$ & $0.930\pm0.153$ & $0.973\pm0.087$ \\
$ 0.144 -  0.288$ & $1.638\pm0.199$ & $1.405\pm0.103$ \\
$ 0.288 -  0.432$ & $2.779\pm0.259$ & $2.561\pm0.139$ \\
$ 0.432 -  0.575$ & $5.025\pm0.347$ & $5.054\pm0.194$ \\
$ 0.575 -  0.719$ & $12.39\pm0.54$ & $11.23\pm0.29$ \\
\hline
  \multicolumn{3}{c}{$\mu^+ \mu^-$} \\ 
\hline
$-0.9 - -0.7$ & $0.115\pm0.101$ & $0.008\pm0.037$ \\
$-0.7 - -0.5$ & $0.057\pm0.044$ & $0.160\pm0.036$  \\
$-0.5 - -0.3$ & $0.139\pm0.066$ & $0.138\pm0.034$ \\
$-0.3 - -0.1$ & $0.080\pm0.049$ & $0.111\pm0.032$  \\
$-0.1 -  0.1$ & $0.396\pm0.127$ & $0.193\pm0.050$ \\
$ 0.1 -  0.3$ & $0.190\pm0.078$ & $0.393\pm0.056$ \\
$ 0.3 -  0.5$ & $0.703\pm0.139$ & $0.383\pm0.056$ \\
$ 0.5 -  0.7$ & $0.427\pm0.111$ & $0.521\pm0.066$ \\
$ 0.7 -  0.9$ & $0.415\pm0.164$ & $0.569\pm0.079$ \\
\hline
  \multicolumn{3}{c}{$\tau^+ \tau^-$} \\
\hline
$-0.9 - -0.7$ & $0.103\pm0.098$ & $0.121\pm0.055$ \\
$-0.7 - -0.5$ & $0.092\pm0.065$ & $0.161\pm0.043$  \\
$-0.5 - -0.3$ & $0.209\pm0.087$ & $0.177\pm0.045$ \\
$-0.3 - -0.1$ & $0.328\pm0.108$ & $0.145\pm0.042$  \\
$-0.1 -  0.1$ & $0.272\pm0.101$ & $0.228\pm0.055$ \\
$ 0.1 -  0.3$ & $0.255\pm0.099$ & $0.328\pm0.061$ \\
$ 0.3 -  0.5$ & $0.375\pm0.119$ & $0.382\pm0.067$ \\
$ 0.5 -  0.7$ & $0.529\pm0.143$ & $0.460\pm0.075$ \\
$ 0.7 -  0.9$ & $1.079\pm0.299$ & $0.450\pm0.113$ 
\end{tabular}
\end{ruledtabular}
\end{table}


\begin{table}[ht]
\caption{
The $e^+e^-$ and hadronic forward-backward asymmetry, as measured at LEP~2.}
\begin{ruledtabular}
\begin{tabular}{ccc}
$\sqrt{s}$ (GeV) & \multicolumn{2}{c}{$A_{\rm FB}$} \\
       & $e^+ e^-$ & $q\bar q$  \\ 
\hline
\multicolumn{3}{c}{ALEPH} \\
\hline
183 & - &   $0.33\pm0.19$ ($b\bar b$) \\
    &   &   $0.95\pm0.27\err{0.11}{0.09}$ ($c\bar c$)  \\
189 & - &   $0.34\pm0.19\pm0.02$ ($b\bar b$)  \\
\hline
\multicolumn{3}{c}{DELPHI} \\
\hline
130.2& $0.81\pm0.06\pm0.015$ & -\\
136.2& $0.89\pm0.04\pm0.013$ & -\\
161.3& $0.82\pm0.04\pm0.012$ & -\\
172.1& $0.81\pm0.04\pm0.01$  & -\\
183&   $0.814\pm0.017$ & - \\
189&   $0.810\pm0.010$ & - \\
192 & $0.831\pm0.024$ & -\\
196 & $0.818\pm0.015$ & -\\
200 & $0.789\pm0.016$ & -\\
202 & $0.829\pm0.020$ & -\\
\hline
\multicolumn{3}{c}{L3} \\
\hline
130   & $0.806\pm0.043\pm0.006$ & - \\
136.1 & $0.879\pm0.039\pm0.006$ & - \\
161.3 & $0.818\pm0.046\pm0.012$ & - \\
172.1 & $0.795\pm0.056\pm0.012$ & - \\
182.7 & $0.778\pm0.021\pm0.004$ & - \\
188.7 & $0.819\pm0.012\pm0.003$ & - \\
\hline
\multicolumn{3}{c}{OPAL} \\
\hline
189 & - & $0.43\pm0.15\pm0.08\;(b\bar b)$  \\
    &   & $0.57\pm0.18\pm0.09\;(c\bar c)$  \\
\end{tabular}
\end{ruledtabular}
\end{table}

\begingroup
\squeezetable
\begin{table}[h]
\caption{Dijet differential cross section $d\sigma/dM$ measured by CDF.}
\begin{ruledtabular}
\begin{tabular}{ccccc}
bin edge & average $M$ & $d\sigma/dM$ & statistical  & systematic \\ 
(GeV$^2$) & (GeV$^2$)  & (pb/GeV$^2$)    & uncertainty  &  uncertainty
  \\ \hline
180 & 188 & 6.07$\times 10^{2}$ &  3.2\,\% & $^{+20}_{-17}$\,\% \\
198 & 207 & 3.42$\times 10^{2}$ &  4.1\,\% & $^{+19}_{-17}$\,\% \\
217 & 228 & 1.81$\times 10^{2}$ &  1.0\,\% & $^{+19}_{-16}$\,\% \\
241 & 252 & 9.81$\times 10^{1}$ &  1.4\,\% & $^{+19}_{-16}$\,\% \\
265 & 277 & 4.98$\times 10^{1}$ &  1.8\,\% & $^{+19}_{-17}$\,\% \\
292 & 305 & 2.78$\times 10^{1}$ &  1.1\,\% & $^{+19}_{-17}$\,\% \\
321 & 335 & 1.43$\times 10^{1}$ &  1.4\,\% & $^{+20}_{-17}$\,\% \\
353 & 368 & 7.41$\times 10^{0}$ &  1.9\,\% & $^{+20}_{-18}$\,\% \\
388 & 405 & 3.83$\times 10^{0}$ &  0.9\,\% & $^{+21}_{-18}$\,\% \\
427 & 446 & 1.89$\times 10^{0}$ &  1.2\,\% & $^{+21}_{-19}$\,\% \\
470 & 491 & 9.07$\times 10^{-1}$ &  1.7\,\% & $^{+22}_{-19}$\,\% \\
517 & 539 & 4.50$\times 10^{-1}$ &  2.3\,\% & $^{+23}_{-20}$\,\% \\
568 & 592 & 1.90$\times 10^{-1}$ &  3.3\,\% & $^{+25}_{-21}$\,\% \\
625 & 652 & 7.42$\times 10^{-2}$ &  5.1\,\% & $^{+26}_{-22}$\,\% \\
688 & 716 & 2.92$\times 10^{-2}$ &  7.7\,\% & $^{+28}_{-23}$\,\% \\
756 & 784 & 1.18$\times 10^{-2}$ & 11\,\%   & $^{+30}_{-25}$\,\% \\
832 & 865 & 3.57$\times 10^{-3}$ & 20\,\%   & $^{+32}_{-26}$\,\% \\
915 & 968 & 9.03$\times 10^{-4}$ & 33\,\%   & $^{+34}_{-28}$\,\%
\end{tabular}
\end{ruledtabular}
\end{table}
\endgroup

\begin{table}[ht]
\caption{Dijet invariant mass distribution measured by 
D\O\ for $|\eta_{\rm jet}| <1$. }
\medskip
\begin{ruledtabular}
\begin{tabular}{cccc}
$M_{jj}$ mass bin (GeV)& 
         \multicolumn{3}{c}{$d^3/dM_{jj} d\eta_1  d\eta_2$ (nb/GeV)} \\
 & $\pm$ stat. error  & syst. low (\%) & syst. high (\%) \\
\hline
$200- 220$  & $(3.78\pm 0.12)\cdot 10^{-2}$ & $-11.4$ & $+11.8$ \\
$220- 240$  & $(2.10\pm 0.09)\cdot 10^{-2}$ & $-11.3$ & $+11.6$ \\
$240- 270$  & $(1.16\pm 0.06)\cdot 10^{-2}$ & $-11.5$ & $+11.7$ \\
$270- 300$  & $(6.18\pm 0.11)\cdot 10^{-3}$ & $-11.5$ & $+12.0$ \\
$300- 320$  & $(3.55\pm 0.11)\cdot 10^{-3}$ & $-11.5$ & $+12.1$ \\
$320- 350$  & $(2.12\pm 0.07)\cdot 10^{-3}$ & $-11.9$ & $+12.3$ \\
$350- 390$  & $(1.18\pm 0.01)\cdot 10^{-3}$ & $-11.1$ & $+11.6$ \\
$390- 430$  & $(5.84\pm 0.09)\cdot 10^{-4}$ & $-11.5$ & $+12.2$ \\
$430- 470$  & $(2.89\pm 0.06)\cdot 10^{-4}$ & $-11.9$ & $+12.9$ \\
$470- 510$  & $(1.64\pm 0.05)\cdot 10^{-4}$ & $-12.4$ & $+13.5$ \\
$510- 550$  & $(8.74\pm 0.34)\cdot 10^{-5}$ & $-12.8$ & $+14.3$ \\
$550- 600$  & $(4.49\pm 0.17)\cdot 10^{-5}$ & $-13.5$ & $+15.3$ \\
$600- 700$  & $(1.73\pm 0.07)\cdot 10^{-5}$ & $-14.9$ & $+17.2$ \\
$700- 800$  & $(4.58\pm 0.38)\cdot 10^{-6}$ & $-17.6$ & $+20.8$ \\
$800- 1400$ & $(2.39\pm 0.35)\cdot 10^{-7}$ & $-23.2$ & $+28.9$ \\
\end{tabular}
\end{ruledtabular}
\end{table}


\begin{table}[ht]
\caption{\label{dj-ang}
Dijet angular distribution in various dijet invariant mass bins 
measured by (a) D\O\ and (b) CDF.  
D\O\ defines $R_\chi \equiv \frac{N(\chi<4)}{N(4<\chi<\chi_{\rm max})}$ while
CDF defines $R_\chi \equiv \frac{N(\chi<2.5)}{N(2.5<\chi<5)}$.
The covariance matrix is given by $V_{ii}=\sigma^2_i({\rm stat}) +
\sigma^2_i({\rm syst.})$ and $V_{ij}=\sigma_i({\rm syst.}) 
\sigma_j({\rm syst.})$. 
}
\label{tablelast}
\begin{ruledtabular}
\begin{tabular}{cccc}
\multicolumn{4}{c}{(a) D\O\ } \\
\hline
\hline
$M_{jj}$ range (GeV)  &  $R_\chi$  & Stat. error  & Syst. error \\
\hline
$260- 425$ & 0.191 & 0.0077 & 0.015 \\
$425- 475$ & 0.202 & 0.0136 & 0.010 \\
$475- 635$ & 0.342 & 0.0085 & 0.018 \\
$>635$    & 0.506 & 0.0324 & 0.028 \\
\hline
\multicolumn{4}{c}{(b) CDF} \\
\hline
\hline
$241- 300$ & 0.678 & 0.012 & 0.018 \\
$300- 400$ & 0.695 & 0.010 & 0.025 \\
$400- 517$ & 0.703 & 0.009 & 0.033 \\
$517- 625$ & 0.738 & 0.023 & 0.054 \\
$>625 $    & 0.732 & 0.046 & 0.103 \\
\end{tabular}
\end{ruledtabular}
\end{table}


\end{document}